# Physarum Inspired Bicycle Lane Network Design in a Congested Mega City

By

**Md. Ahsan Habib**
**Roll: 1507082**

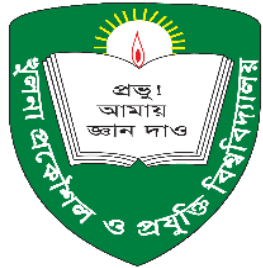

**Department of Computer Science and Engineering**

**Khulna University of Engineering & Technology**

**Khulna 9203, Bangladesh**

**March 2020**

# Certification

The thesis titled "**Physarum Inspired Bicycle Lane Network Design in a Congested Mega City**" submitted by Md. Ahsan Habib, Roll No: 1507082, Academic Year: 2018-19, for partial fulfillment of the requirements for the degree of "Bachelor of Science in Computer Science and Engineering".

**Supervisor**

---

Dr. Muhammad Aminul Haque Akhand

Professor

Dept. of Computer Science and Engineering

Khulna University of Engineering & Technology

Khulna, Bangladesh.

# Acknowledgements

First and foremost, I must sense grateful to and wish to acknowledge my insightful indebtedness to Dr. Muhammad Aminul Haque Akhand, Professor of Department of Computer Science and Engineering and the supervisor of the thesis. His unfathomable knowledge in this field influenced me to carry out this thesis up to this point. His endless endurance, scholarly guidance, continual encouragement, constant and lively supervision, constructive criticism, priceless suggestion made it possible to come up to this phase. Without his inspiring, enthusiasm and encouragement, this work could not be completed.

Last, but by no means least, I thank Allah for the talents and abilities I was given that made it possible to undertake this thesis.

# Abstract

Mobility is a key factor in urban life and transport network plays a vital role in mobility. Worse transport network having less mobility is one of the key reasons to decline the living standard in any unplanned mega city. Transport mobility enhancement in an unplanned mega city is always challenging due to various constraints including complex design and high cost involvement. The aim of this thesis is to enhance transport mobility in a megacity introducing a bicycle lane. To design the bicycle lane natural Physarum, brainless single celled multi-nucleated protist, is studied and modified for better optimization. Recently Physarum inspired techniques are drawn significant attention to the construction of effective networks. Exiting Physarum inspired models effectively and efficiently solves different problems including transport network design and modification and implication for bicycle lane is the unique contribution of this study. Central area of Dhaka, the capital city of Bangladesh, is considered to analyze and design the bicycle lane network bypassing primary roads.

# Contents

# List of Tables

# List of Figures

# CHAPTER 1

# Introduction

A transport network can be described as a collection of linear features that permits either vehicular movement or flow of some commodity. The characteristics of Physarum can be used to design a transportation network. This chapter discusses the background study of network design, objectives, and organization of the thesis.

## 1.1 Overview of Transport Network in a Mega City

Mobility, a key factor in planning and designing urban transport, is a fundamental part of human beings [1]. For mobility purposes, people can use both motorized and non-motorized vehicles within a city. Motorized vehicles like buses, cars, motorbikes, cycles, etc. are hazardous in many kinds [2]. Non-motorized transports involve walking and cycling as well as variants such as small-wheeled transportation like skates, skateboards, push scooters and hand carts [3]. Nowadays, non-motorized mobility is trendy [4].

The idea of mega city emerged to characterize the world's largest metropolitan agglomerations at the end of the 20th century. In the 1970s, only two mega cities had over ten million residents. Currently, 9.9% of the urban population globally resides in 23 megacities. It is projected that in 2025, the number will increase to 37 if 13.6% of global urban population are to be accommodated [5].

A transportation network is the formation of a spatial network that enables vehicle movement or the flow of some commodities. A vast network of rail, subways and bus lines passes through well-organized mega cities. There are also exist special footways and networks for cycling routes.

## 1.2 Motivation

In a well-planned and well-organized mega city, both footways and roads are available for mobility purpose but in case of an unplanned and unorganized mega city, both footpaths and roads are hardly available. In some cases, footpaths are snatched by hawkers and street vendors. The public transport system is defined by far a lack of people's desired travel needs in terms of mobility, reliability, convenience, pace and safety. In fact, some transports like buses are considered unreliable and time consuming to reach their destinations. The Texas Transportation Institute reported a delay of 3.6 billion vehicle-hours in the 75 biggest metropolitan regions in 2000, culminating in 5.7 billion U.S. gallons (21.6 billion liters) of waste fuel and a loss of productivity of $67.5 billion or around 0.7 percent of GDP.

Bicycling or walking activity may help increase blood flow, release endorphins, and decrease overall stress. It can even help to improve mental health and energy by tracking 30 minutes of bicycling or walking a day[6]–[9]. Efficient and effective bicycle lane network design in an unplanned and unorganized mega-city can minimize total travel time, fuel usage, costs, carbon dioxide ($CO_2$) emission, etc. Physarum polycephalum is multi-headed, brainless, a giant multi-nucleated, single-celled protist that can solve different complex problems. Physarum networks are believed to have achieved a good balance between cost, efficiency, and resilience.

## 1.3 Objectives of the Thesis

In the case of an unorganized and unplanned mega city, where the transportation network is congested and unplanned and there are hardly any footpaths available and no further transport facilities can be expanded. So there are some huge problems in those cities like traffic jam, noise pollution, air pollution, $CO_2$ emission, etc. With a planned lane network with non-motorized vehicles nearly all of this problem can be solved. Since it is not feasible to completely rebuild the transportation network and infrastructure of a mega city but possible to transform mega city towards a green city. The objective of this study is given below:

- Study of Physarum
- Physarum related paper study
- Network design
- Bicycle lane network in mega city

## 1.4 Organization of the Thesis

The main attraction of this thesis is to present a modified Physarum inspired technique to construct bicycle lane network design. The thesis has five chapters. An introduction to network design and Physarum has been given in Chapter 1. Chapter-wise overviews of the rest of the thesis are as follows:

**Chapter 2:** Describes the literature review that includes a brief description of Physarum with its properties and previous related work to Physarum inspired network design.

**Chapter 3:** Explains the proposed modified Physarum Inspired Bicycle Lane Design in an Unplanned Mega City in detail.

**Chapter 4:** Reports the experimental result of modified Physarum Inspired Bicycle Lane Design. Also, in this chapter, a case study of Dhaka is demonstrated. Finally,

**Chapter 5:** This chapter is for the conclusions of this thesis together with the outline of future directions of research opened by this work.

# CHAPTER 2

# Physarum Inspired Network Design

Physarum polycephalum is a brainless amoeboid organism. Physarum-inspired network design model has demonstrated extraordinary skill in designing effective networks. In this chapter firstly, we discuss the Physarum polycephalum, secondly the Physarum-based network design and lastly, the existing network design inspired by Physarum.

## 2.1 Physarum and its Properties

Physarum polycephalum, accurately the 'many-headed' slime mold, is a gigantic multi-nucleated but single-celled protist [10]. The slime mold Physarum polycephalum creates a form of spatial memory by avoiding areas it has previously explored to navigate in a complex environment[11]. Recently, Physarum polycephalum (true slime mold) has arisen as a fascinating illustration of biological computation through morphogenesis[12]. Although it is a single-cell organism, studies have shown that the Physarum can overcome different minimum cost flow problems through its growth process[12]. In the following Fig. 2.1, an example of the Physarum polycephalum is shown.

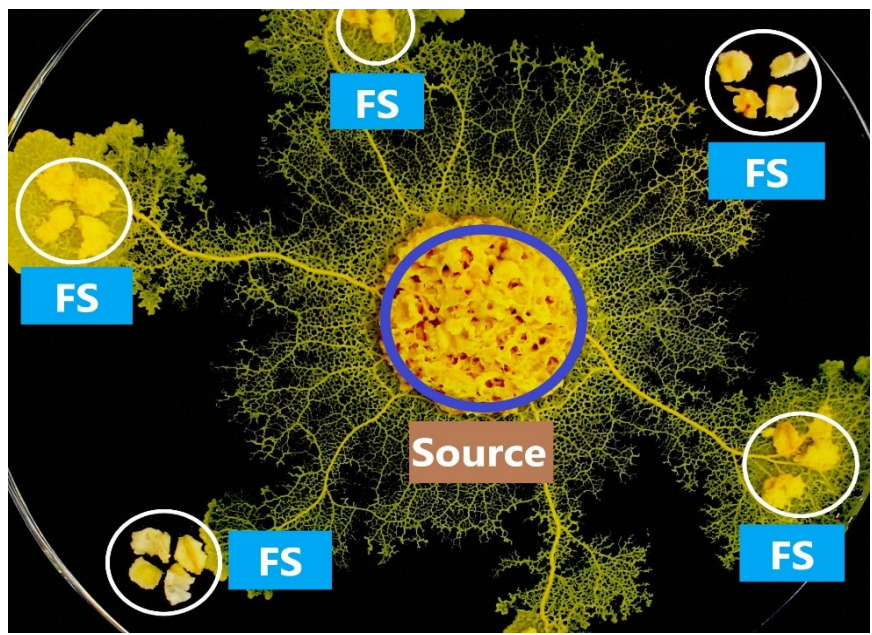

Figure 2.1: Physarum polycephalum. [61].

Here Physarum polycephalum is shown to grow up the network towards the FSs from the source. (FS = Food Source)

## 2.2 Network Design Inspired on Physarum

The intelligent behavior of slime mold was first observed by Nakagaki et al. in 2000[13]. In previous biological experiments, Physarum-inspired network model has exhibited an extraordinary intelligence to build efficient networks to connect multiple food sources. Physarum networks are believed to have achieved a good balance between cost, efficiency, and resilience. For instance, Physarum constructed networks with comparable qualities to those of the Tokyo rail system in a renowned experiment performed by Tero et al. in 2010[14].

They developed a mathematical model for adaptive network construction to emulate the behavior of Physarum which is based on feedback loops between the thickness of each tube and internal protoplasmic flow in which high rates of streaming stimulate an increase in tube diameter, whereas tubes tend to decline at low flow rates. The edges represent plasmodial tubes in which protoplasm flows, and nodes are junctions between tubes. They consider the pressure at nodes $i$ and $j$ are $P_i$ and $P_j$, respectively, and the two nodes are connected by a cylinder of length $L_{ij}$ and radius $r_{ij}$. They assume that the flow is laminar and follows the Hagen-Poiseuille equation, the flux through the tube is,

$$Q_{ij} = \frac{\pi r_{ij}^4 (P_i - P_j)}{8\varepsilon L_{ij}} = \frac{D_{ij}(P_i - P_j)}{L_{ij}}, \qquad (2.1)$$

here $\varepsilon$ is the viscosity of the fluid, and $D_{ij} = \frac{\pi r_{ij}^4}{8\varepsilon}$ is a measure of the conductivity of the tube. As the length $L_{ij}$ is a constant, the behavior of the network is described by the conductivities of the edges.

The constrains must be maintained,

$\sum_j Q_{1j} = I_0,$      For source node-1

$\sum_j Q_{2j} = -I_0,$      For sink node-2

$\sum_j Q_{ij} = 0,$      Inflow and outflow must be conserved

To accommodate the adaptive behavior of the plasmodium, the conductivity of each tube evolves according to $\frac{dD_{ij}}{dt} = f(|Q_{ij}|) - D_{ij}$, where $f(|Q_{ij}|)$ describes the expansion of tubes in response to the flux and $D_{ij}$ represents the rate of tube constriction, so the tubes will gradually disappear.

The functional $f(|Q|) = \frac{|Q|^\gamma}{(1+|Q|^\gamma)}$ which describes a sigmoidal response where $\gamma$ is a parameter that controls the nonlinearity of feedback ($\gamma > 0$).

## 2.3 Review of Existing Physarum Inspired Works

Physarum can successfully overcome many problems in real life even more complicated problems. In this section initially, we discuss and summarize about the Physarum inspired network design techniques, and then discuss about other optimization problems solved using Physarum inspired methods.

## 2.3.1 Transpiration Network Design

To link several food points Physarum can build high-quality networks. A mathematical model of the adaptive dynamics of a transport network of the true slime mold that shows path-finding behavior in a maze is developed in 2007 by Tero et al. [15]. In 2010, Physarum developed networks shows similar qualities to the Tokyo rail system in a famous experiment conducted by Tero et al. [14]. Since then, Physarum inspired other real-world transport networks, such as Iberian motorways [16] and Mexican Federal highways [17] have also been constructed. Adamatzky et al. [18] develops a model to construct networks on major urban areas of China. Becker et al. [19] developed in 2011 a fault tolerant connection networks for the Tokyo rail system using an agent based simulation of Physarum polycephalum. Physarum-inspired cellular automaton (CA)-based network designing model was developed by Tsompanas et al. [20] inspired by Slime Mould. Zhang et al. [21] recently proposed a method to solve the problem of network design in supply chain for multiple source nodes and multiple sink nodes. Physarum is excellent at doing other network design [22]. Here we summarize various works of network construction using Physarum inspired technique in the following Table 2.1.

Table 2.1: Network construction using Physarum.

| Authors & Year | Title of Paper | Contribution |
|---|---|---|
| Tero et al., 2007 [15] | A mathematical model for adaptive transport network in path finding by true slime mold | Model for adaptive transport network in Path-finding in a maze |
| Tero et al., 2010 [14] | Rules for Biologically Inspired Adaptive Network Design | Tokyo Rail Network construction |
| Adamatzky et al., 2011 [16] | Rebuilding Iberian motorways with slime mould | Iberian motorway network construction |
| Adamatzky et al., 2011 [17] | Approximating Mexican highways with slime mould | Mexican Federal highway network construction |
| Adamatzky et al., 2013 [18] | Slime mould imitates transport networks in China | Slime mould protoplasmic networks on major urban areas of China |
| Becker et al., 2011 [19] | Design of fault tolerant networks with agent-based simulation of Physarum polycephalum | Construction of fault tolerant connection networks for the Tokyo rail system using an agent based simulation of Physarum polycephalum |
| Tsompanas et al., 2015 [20] | Evolving Transport Networks With Cellular Automata Models Inspired by Slime Mould | Physarum-inspired cellular automaton (CA)-based network designing model |
| Zhang et al. 2016 [21] | A Physarum-inspired approach to supply chain network design | Supply chain network design |

## 2.3.2 Other Optimization Task

Nakagaki et al. in 2000 [13] observed that Physarum productively found the shortest path between two selected points in a maze. In addition, the Physarum can solve many other famous problems like the shortest paths [23]–[25], towers of Hanoi problem [26] and minimum risk problem [27]. Physarum can effectively solve many other complex problems in the real world like traveling salesman problem [28]–[30], population migration [31], etc. Logic gates design and boolean operations can be performed by a slime mold network [32], [33]. Chaining these logic gates together can enable a slime mold computer to perform binary computation operations. Physarum works very well in logical computing as well[34]–[38]. Identifying critical components [39], [40] and many other problems [41], [42] are effectively and efficiently solved through Physarum bio-inspired technique. Most interestingly, many other studies have shown that Physarum's tubular topologies often mimic those of complex mathematical networks [43], [44] like the Steiner tree problems [45]–[49].

Studies with Physarum related works showed that the organism can solve many complex real-life problems efficiently and effectively, particularly in the sense of network design. This can be applied with some changes for designing the bicycle lane network. In this work bicycle lane network is planned using local lanes in congested mega city.

# CHAPTER 3

# Physarum Inspired Bicycle Lane Design in an Unplanned Mega City

An unplanned mega city suffers various problems including transporation and mobility. Transport mobility enhancement in an unplanned mega city is always challenging due to various constraints including complex design and high cost involvement. In this thesis, we try to increase the mobility in an unplanned mega city Dhaka. In this chapter, problems of an unplanned mega city Dhaka are addressed firstly, then challenges in transformation an unplanned megacity to green city, and finally, the importance of bicycle lane in an unplanned mega city and bicycle lane network design in an unplanned mega city are discussed.

## 3.1 Mobility Problem in an Unplanned Mega City: Dhaka as a Case Study

This thesis aim is to enhance transport mobility in an unplanned mega city introducing a bicycle lane. In this section initially, we discuss the history and overview of Dhaka city, then the transportation crisis in Dhaka city and finally, the effect of transportation crisis on other problems.

### 3.1.1 History and Overview of Dhaka City

It is mentioned that the concept of a Mega City originated at the end of the 20th century to describe the largest city in the world. Although, literature has little disagreement about the population threshold used as a megacity concept, the UN (2003) defines most precisely: a conurbation of ten million or more inhabitants is a megacity which has now been widely accepted [5].

Since 1971, Dhaka has experienced incredible growth and rapid growth. It is one of the world's only seven cities with a population of over 2.4 percent between 1975 and 2005 (UN 2006). In 2011, it was one of the world's top ten mega cities. The developments have unfortunately happened unplanned, especially since the 1990s [5]. The word

Dhaka is nowadays mentioned regularly in the most unlivable cities. Dhaka was the world's fastest-growing town between 1950 and 2000 [50]. While population growth has declined recently, it is still the second-largest growth mega city in the world [50].

### 3.1.2 Transportation Crisis in Dhaka City

The mega city has neither efficient public transport nor mass transit [50]. It is probably the world's only mega-city without efficient public transit and public transit [50]. Dhaka has a poorly developed transport system with 200 km of main roads and about 260 km (too few) secondary and collector roads, in addition to 250 km of narrow roads (approximately) [50]. There are many incomplete critical connections in the road network and several regions have insufficient connectivity to the network [50]. Separate bicycling lanes and footpaths are barely available in the city, which enhance the mobility crisis.

There was a time when traffic congestion was only suffered by commuters on the main streets of the city, but now it starts right from the door. Traffic jam has turned into nightmares for daily trips. According to a World Bank report, the average traffic speed in Dhaka has dropped from 21 kilometers per hour (kmph) to 7 kilometers per hour in the last 10 years, and by 2035 the speed could drop to 4 kilometers per hour, which is slower than the walking speed [55]. Another study commissioned by the BRAC Institute of Government and Development indicates that traffic congestion in Dhaka consumes about 5 million working hours a day and costs the country $11.4 billion a year [55]. The financial loss is a measure of the time lost in traffic congestion and the extra hours expended on cars.

It should be noted that there is no adequate and proper routing of our public transport system. In 2016, According to the BRTA, 20,304 new cars were introduced to Dhaka's traffic, which means more than 55 new cars hit the streets every day [55]. As the number of cars increases, there is also an increasing demand for parking space. Unfortunately, however, the parking space in our city is quite inadequate. Many vehicles on the streets are stored. Many buses and trucks are parked on the streets on a regular basis [55].

According to the Dhaka Metropolitan Police (DMP) Traffic Department, traffic jams have become intolerable in some urban areas over the past few days, including Mirpur-12 to Mirpur-10 crossing, Rokeya Sarani, Gulshan, Banani, Badda, Moghbazar, Eskaton, Tejgaon, Airport Road, and Uttara, for a number of reasons, including the ongoing Dhaka International Trade Fair, the building of underground trains and the increase in private transport[56]. Urban analyst and former chairman of UGC Prof Nazrul Islam said traffic jams are gradually deteriorating due to an increase in urban population and the number of small vehicles and lack of effective control measures [56]. "We have built over half dozens of flyovers, but it is not a solution to solve the problem. We will not be able to reduce traffic jams without increasing public transport and ensuring better traffic management", he observed [56]. Transport and urban experts believe that the government should take practical steps to ensure effective mass transportation, restore transportation efficiency, decrease the use of private and small cars, replace micro-buses and mini-busses with single-decker, double-decker, and articulated buses, and extend the city to dramatically alleviate traffic jams without spending huge money [56]. The experts also said that railways and waterways can also be used effectively to relieve road traffic pressure and facilitate trouble-free transport services for the commuters [56].

Figure 3.1 illustrates the traffic jam in the city. Here, three routes are available from Farmgate to the University of Dhaka. During driving mode, it takes around 15 minutes at 06:00 a.m., 18-35 minutes at 10:00 a.m., and 18-40 minutes at 5:00 p.m. on average. On the other side, it takes an average of 45-50 minutes in walking mode. We note that the speed of driving is slightly higher than that of walking. Not only in some areas, but throughout the city, it's the case.

The number of automobiles has been increasing in Dhaka city at the rate of at least 10 percent annually, which has been contributing to environmental pollution on the one hand and traffic congestion on the other. This transportation problem enhances other problems like air pollution, noise pollution, fuel consumption, $CO_2$ emission, worst in road conditions, etc.

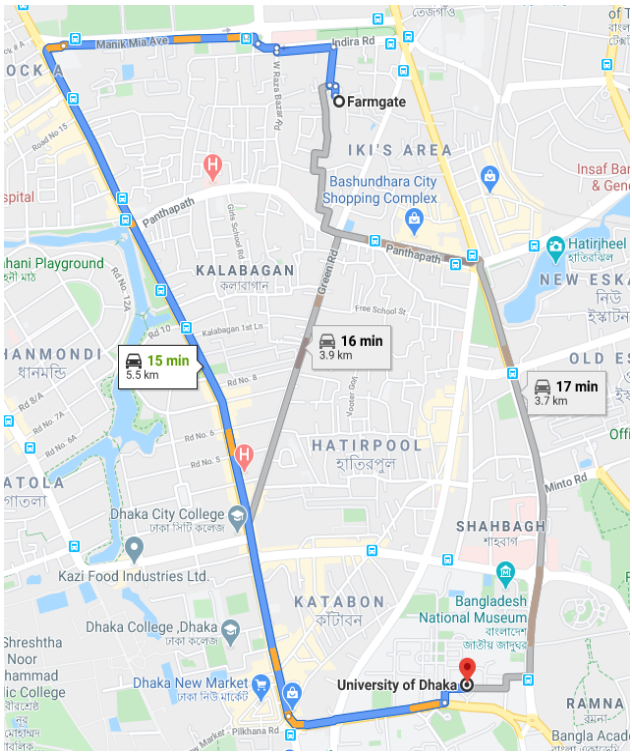

(A) At 06.00AM

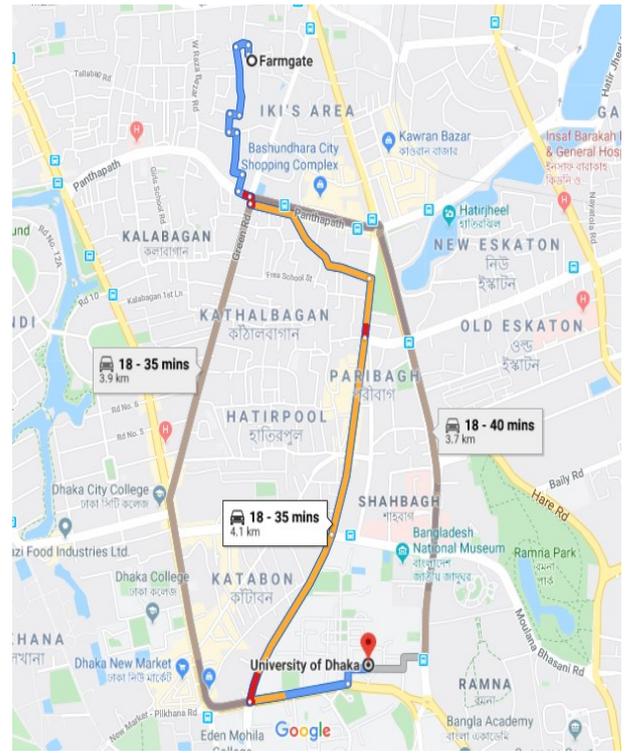

(B) At 10.00AM

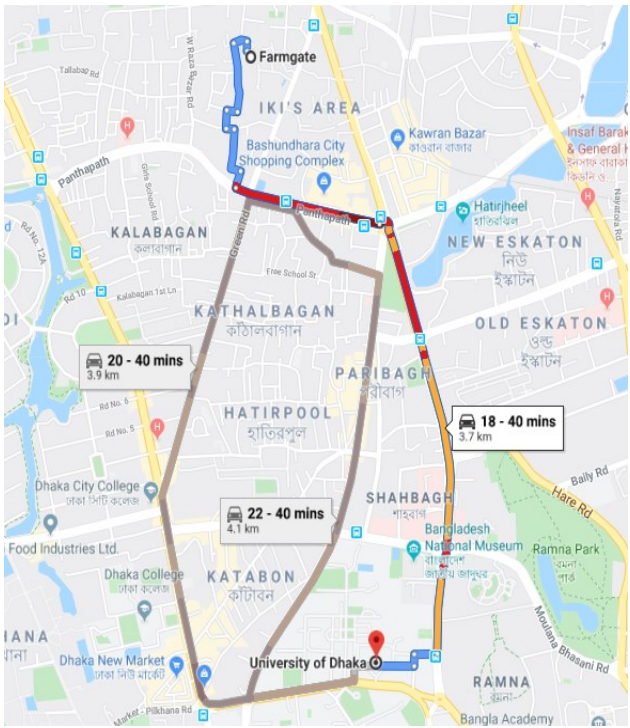

(C) At 5.00PM

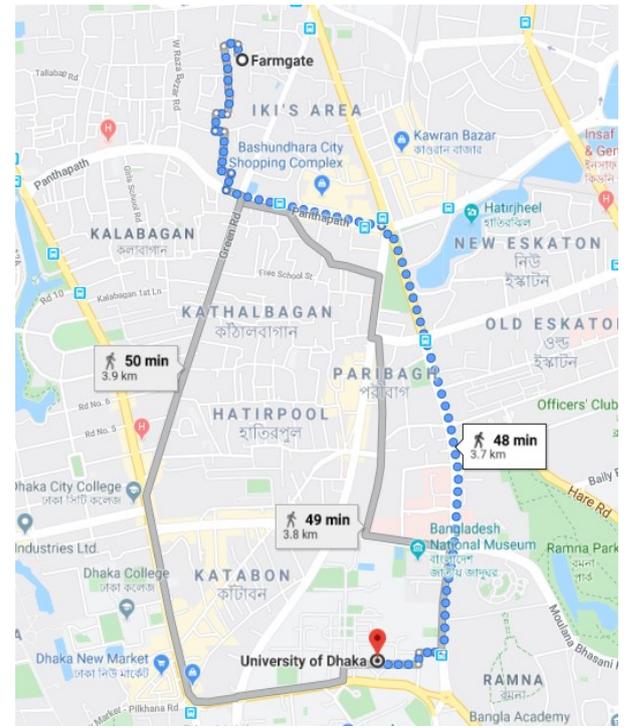

(D) Walking mode

Figure 3.1: Farmgate to University of Dhaka routes and time needed in driving mode. (A) At 06.00am. (B) At 10.00am. (C) At 5.00pm. (D) Time needed in walking mode.

### 3.1.3 Effect of Transportation Crisis on Other Problems

Transportation crisis affects the environment badly. It may affect the air pollution, noise pollution, fuel consumption, etc. According to the Department of Environment (DoE), the standard value of the Air Quality Index (AQI) is 50 represents good air quality with little potential to affect public health [51]. But, according to AirVisual information, Dhaka the capital city of Bangladesh has been ranked the worst in the Air Quality Index

 (AQI) valued 309, which is hazardous and would trigger health warnings of emergency conditions [51]. The entire population is more likely to be affected by the enormous number of diseases like nausea, asthma, high blood pressure, heart disease, and cancer. It also impacts the respiratory tract severely and causing irritation. Children's cognitive faculty will be adversely affected by lead exposure, which can also distress the central nervous system, causing hypertension and renal injury. In the last two months, the capitalist has been enjoying just nineteen hours of good air [51]. Diesel-run vehicles account for more than 80 percent of the air pollution in Dhaka as most of them fail to comply with the approved emission standard, said a recently published survey report [52].

In Dhaka, the average sound level is between 80dB and 110dB in prime areas such as Farmgate, Karwan Bazar, Shahbagh, Gabtoli, and Mohakhali Bus Terminal, says the study report [53]. According to the World Health Organization (WHO), this is almost twice the maximum noise level that can be tolerated by humans – 60dB – without suffering a gradual loss of hearing [53]. According to a recent study conducted by WHO at 45 locations of Dhaka city, most of the traffic points and many of the industrial, residential, commercial, silent and mixed areas are suffering noises exceeding the standard limits of Bangladesh [54]. WHO has also identified several areas as severe red, moderate red, mild red and green zones in terms of noise pollution in Dhaka city [54]. Around 11.7% of the population in Bangladesh have lost their hearing due to noise pollution, says the Development of Environment (DoE) study, which was conducted in 2017 [53]. The major sources of noise pollution in urban areas are traffic and loud horns. The DoE found that in Dhaka, 500-1,000 vehicles honk at the same time when stuck in traffic [53]. Around 5% of the world population is facing several kinds of health hazards due to complexities related to noise pollution, According to the WHO [53].

There is a scarcity of natural gas and petroleum in Bangladesh also. Gas supplies meet 56% of domestic energy demand [57]. Bangladesh has a very limited energy reserve; small amounts of oil, coal and countable natural gas reserves [58]. The country is a net importer of crude oil and petroleum products [57].

## 3.2 Importance of Bicycle Lane in Mega City

In more ways than one, driving a bicycle has a positive impact on the environment. They are also less expensive than other forms of transportation and environment-friendly. Bicycles are considered zero-emission vehicles i.e. they do not release any carbon emissions. Bicycles, as vehicles with zero emissions, do not contribute to air pollution. People can have moderate fresh air. They do not contribute to sound pollution. When bicycles are used as a consistent form of travel by a large percentage of the population in a particular area especially in an urban area, there is a great relief on road traffic conditions. Bicycles also have the effect of alleviating parking difficulties in urban areas, because they simply take up so much less space than cars. Low physical activity or Physical inactivity is recognized as one of the country's leading risk factors and the fourth leading cause of deaths due to non-communicable disease (NCDs) worldwide - cardiovascular diseases, chronic lung diseases, heart disease, stroke, diabetes and cancers - and each year contributes to over three million preventable deaths [59]. There may be some physical exercise every day by using bi-cycle. Bicycles also offer more freedom of movement without time constraints, crowded and unpleasant conditions and, if desired, the ability to travel alone. So, people can have eco-friendly Travel. Bicycles are lighter and usually cause less damage to the roads than others. This will reduce the number of injuries. So the area would be environment-friendly. Most of the well-organized mega city criteria are met by using bicycles.

## 3.3 Challenges to Increase Mobility in Dhaka City

In the case of an unplanned or unorganized city, one of the big issues is that road conditions are not good enough and the cycling lanes & footways are hardly available which is the major cause of worst traffic. On the other hand, noise pollution and traffic congestion are troubling and there is a huge $CO_2$ gas emission occurs.

Between the well-organized city and unplanned city, there is a huge gap in road conditions, footways & cycling lanes, air quality, noise pollution, and traffic congestion. There may have some parameters to increase transport mobility in the city like the construction of separate roads, underground roads, railways, etc. But the construction of those parameters is not a feasible solution because of huge budgets and spaces. There are two alternative low-cost solutions exist, the first one is to make ready the footpaths for routing purpose and the next one is to introduce local lanes with bicycles as vehicles for moving around the city. But the first one is not possible because most of the time, street vendors and hawkers snatch up footways and in some case footways not exist. The next solution is feasible and effective as bicycles have several environmental benefits.

For this purpose, we have to plan a network and always try to use local lanes for routing through one place to another place, if it is not feasible to use local roads for some cases, we will use main roads and will always try to minimize the use of main roads. There exist some constraints to be handled to plan network that we cannot access all possible roads like VIP roads, heavy traffic roads, etc. On the other hand, it is almost impossible to plant more trees to improve air quality and reduce $CO_2$. And most of the noise pollution and traffic congestion is caused by motor vehicle use.

## 3.4 Bicycle Lane Design in an Unplanned Mega City

Different approaches have been presented over the past decades to design networks. It is possible to split the solutions into two categories: exact solutions and heuristic solutions. Exact approaches can treat Network Design Problem in a rigorous way which is inefficient when dealing with real-world large-scale networks. And, an approximate yet efficient approach is provided by heuristic approaches, more popular than exact approaches, that have emerged in recent decades which can tackle large-scale real-world problems. Without using an exact and heuristic approach, here we present the Physarum-inspired technique which takes into account the constraints to construct the bicycle lane network. Basically, we always try to use local lanes for routing through one place to another place, if it is not feasible to use local roads for some cases, we will use main roads and will always try to minimize the use of main roads.

Compared to previous studies it is noted that the network has only one direction between two nodes, so the stream is only flowing from one node to another, but is never flowing in the opposite direction. However, most roads have the features of double-way traffic in real traffic networks as demonstrated in Fig. 3.2. There is a clear distinction between opposite directions, where flows do not interfere in two directions opposite. Apparently, in the traffic network shown in Fig. 3.2, the initial approach influenced by the Physarum cannot be applied.

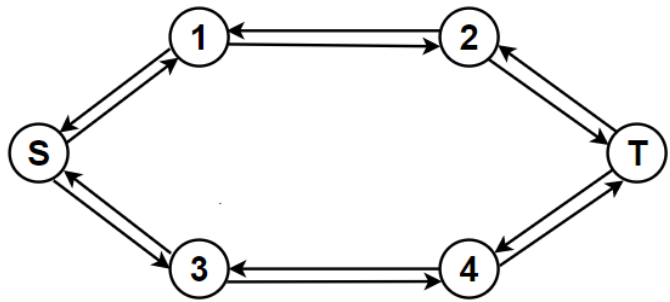

Figure 3.2: Real traffic network.

Here in the following, we discuss the modified Physarum inspired lane design technique.

Given a graph $G = (N, E)$, where

$N$ denotes a set of $n$ cities,

$E$ represents a set of $m$ connections or linkages.

There is a protoplasmic flow in each link of this model. The two terminals of the link represent two locations of the specified area. One terminal is called the source node, and the other terminal is called the sink node. Protoplasmic flows from the source node into the network and from the sink node out of the network. At each city there is pressure and the amount of flux in each edge is proportional to the difference in pressure between the two terminals of this edge. Specifically, the flux $Q_{ij}$ in edge (i,j) is given by the modified Hagen-Poiseuille equation below.

$$Q_{ij} = \frac{D_{ij}}{c_{ij}}(P_i + P_j) \tag{3.1}$$

$$D_{ij} = \frac{\pi r_{ij}^4}{8\varepsilon} \tag{3.2}$$

In the above equation, $D_{ij}$ is the conductivity of the linkage, $c_{ij}/L_{ij}$ is the length of the edge, $P_i$ and $P_j$ are the pressure of the vertices $i$ and $j$, $r_{ij}$ is the radius of the edge, $\varepsilon$ (epsilon) is the coefficient of viscosity. In the case of conductivity($D_{ij}$), which is linkage specific, we are using a fixed conductivity value for all the linkages for simplicity. The length($c_{ij}/L_{ij}$) is not the direct length from $i$ city to $j$ city rather we consider all possible path length with no use or hardly use of main roads and we calculate pressure of each city based on the amount of population in that city. As we use initial fixed conductivity value, $r_{ij}$ is considered to be the same for all connections. Eq. (3.2) indicates that the tubular thickness($r_{ij}$) of Physarum increases with the conductivity of the tube. Therefore, the conductivity update formula can explain the change in tubular thickness of Physarum as follows,

$$\frac{d}{dt}D_{ij} = f(|Q_{ij}|) - \mu D_{ij}, \tag{3.3}$$

here $f(|Q_{ij}|)$ is an increasing function, $\mu$ is a positive constant. In our case, we considered $f(|Q_{ij}|) = |Q_{ij}|$ for simplicity. The equation of conductivity update suggests that conductivity tends to increase with large flux edges. Consequently, the equation of conductivity update reflects the above physiological mechanism. We must first calculate the pressures to calculate the flux and update the edge conductivities. The pressures can be determined using the Poisson equation network below by considering the flux conservation law at each vertex,

$$\sum_{i \in V(j)} \frac{D_{ij}}{c_{ij}}(P_i + P_j) = \begin{cases} -I_0, & j = source \\ +I_0, & j = sink \\ 0, & otherwise \end{cases}, \tag{3.4}$$

here $V(j)$ is the set of vertices linked to vertex $j$, $I_0$ is the amount of flux flowing into and out of the node of the source.

Let the pressure at the sink node be 0, and give an initial value to each edge conductivity, then use Eq. (3.4) to measure the other pressures. After that, we can calculate the amount of flux in each edge using Eq. (3.1), and we can change the conductivity of each edge using Eq. (3.2). According to an edge conductivity threshold value, edges with conductivity lower than this value are cut off from the network.

Let's consider a simple small network consist of only eleven points picked from Dhaka City. Physarum always finds the optimal route network among the eleven nodes and is believed to have achieved a good balance between cost, efficiency, and resilience. Here are the illustrations in the following which are the generalized design using modified Physarum inspired technique. This technique can be applied any real-life traffic network design. In this work, the technique is applied both prominent area centered with Motijheel and entire Dhaka city.

In Fig. 3.2, (A) there are 11 node points between Farmgate (node point 2) and the University of Dhaka (node point 10) existing network with both main and local roads. Here, (B) – (E) shows that the network growing process with time (t = iteration) i.e. (B) Network at 10th iteration; (C) Network at 20th iteration; (D) Network at 50th iteration; (E) Final network for 11 nodes. For those cases, we always try to avoid main roads to grow up the networks. And the Fig. 3.3 depicts the possible road network design if all main roads are available.

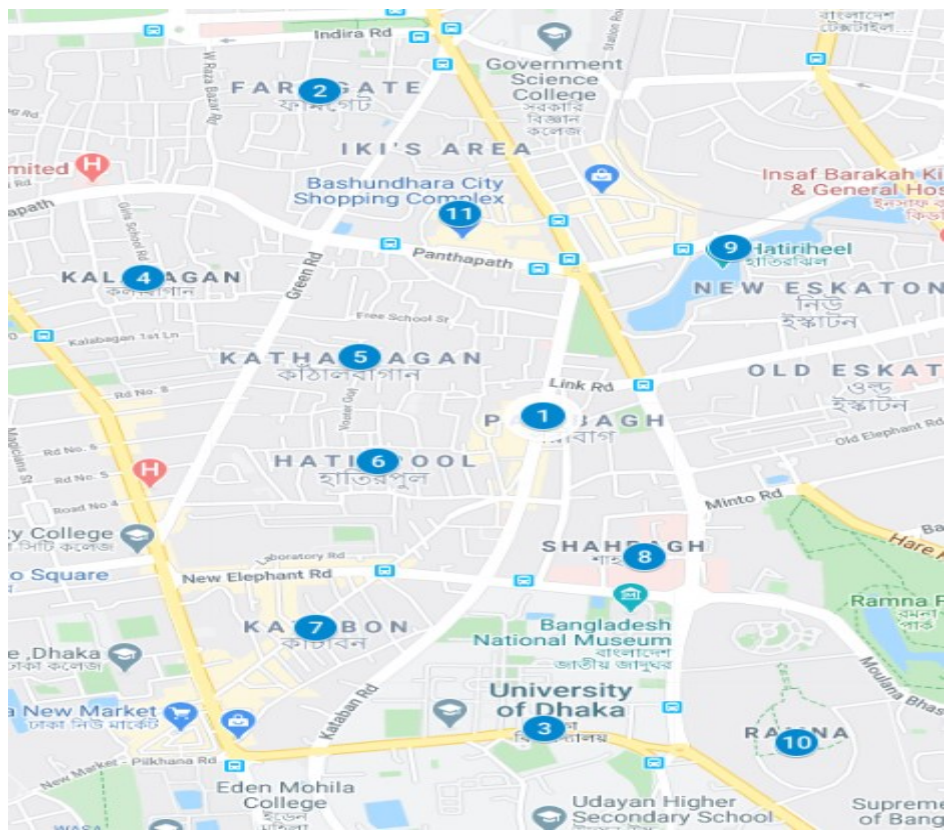

Figure 3.3: Sample Region with 11 points.

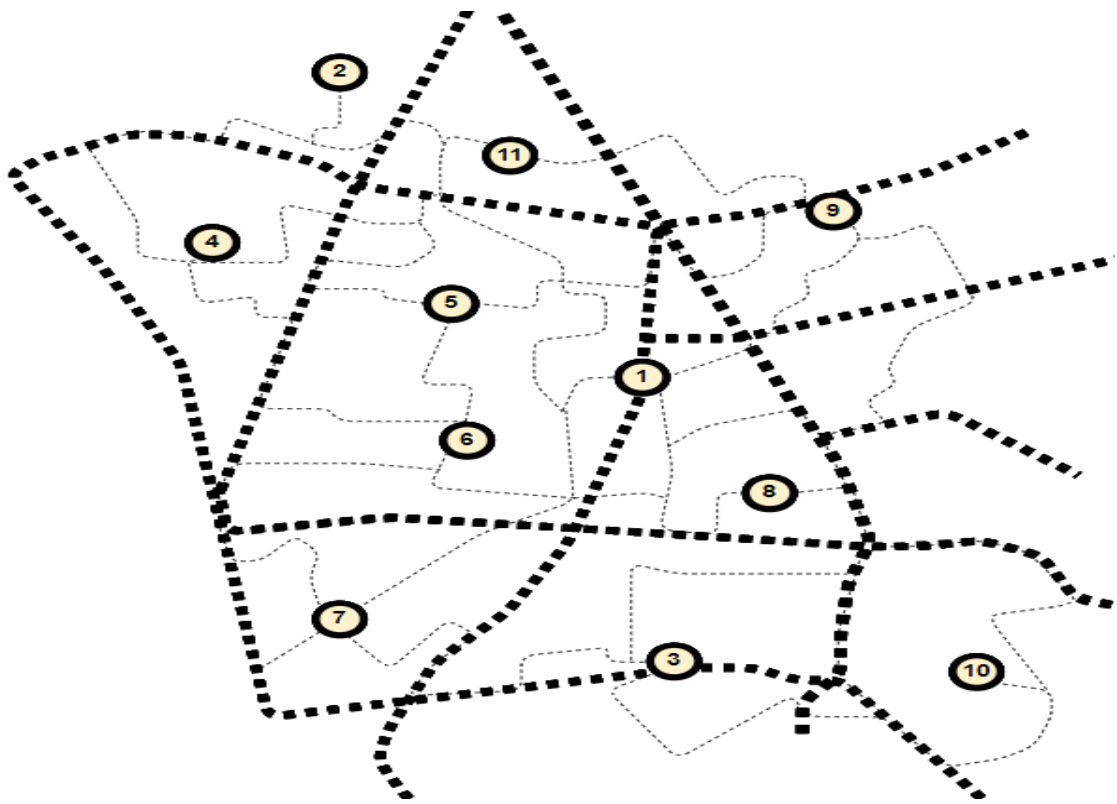

(A) 11 node points (t = 0)

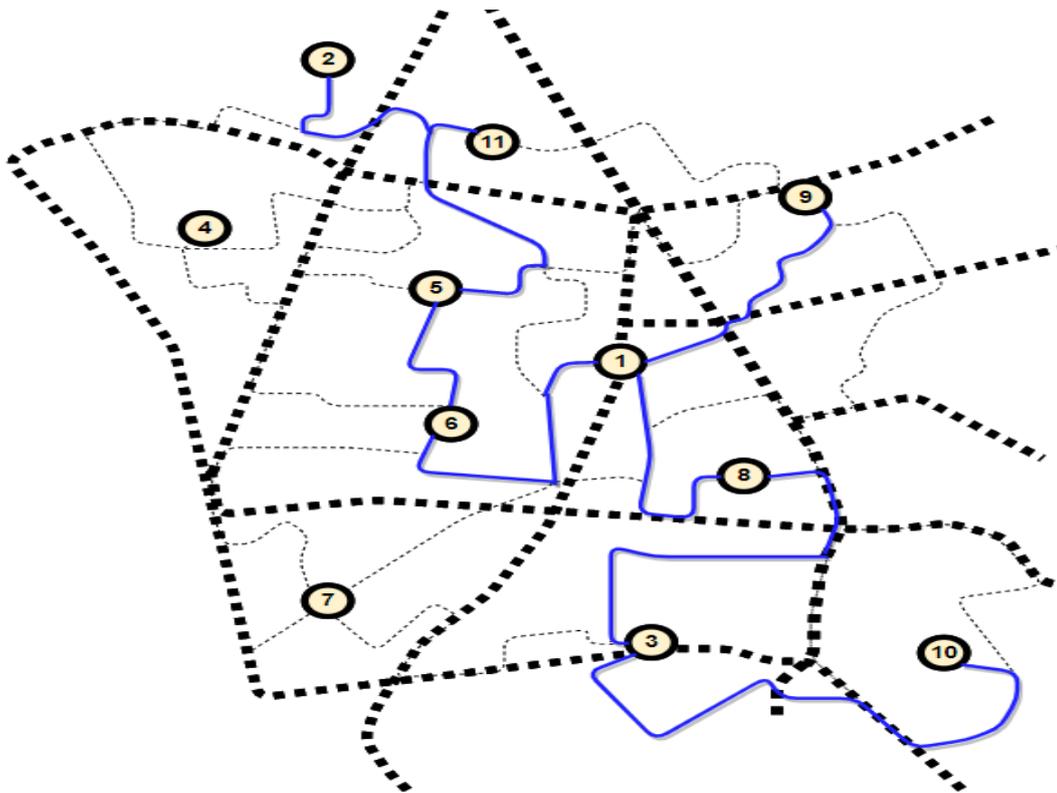

(B) At t = 10

Figure 3.4: Physarum inspired network design of 11 nodes. (A) 11 node points. The network is expanding with *t*. (B) At *t* = 10.

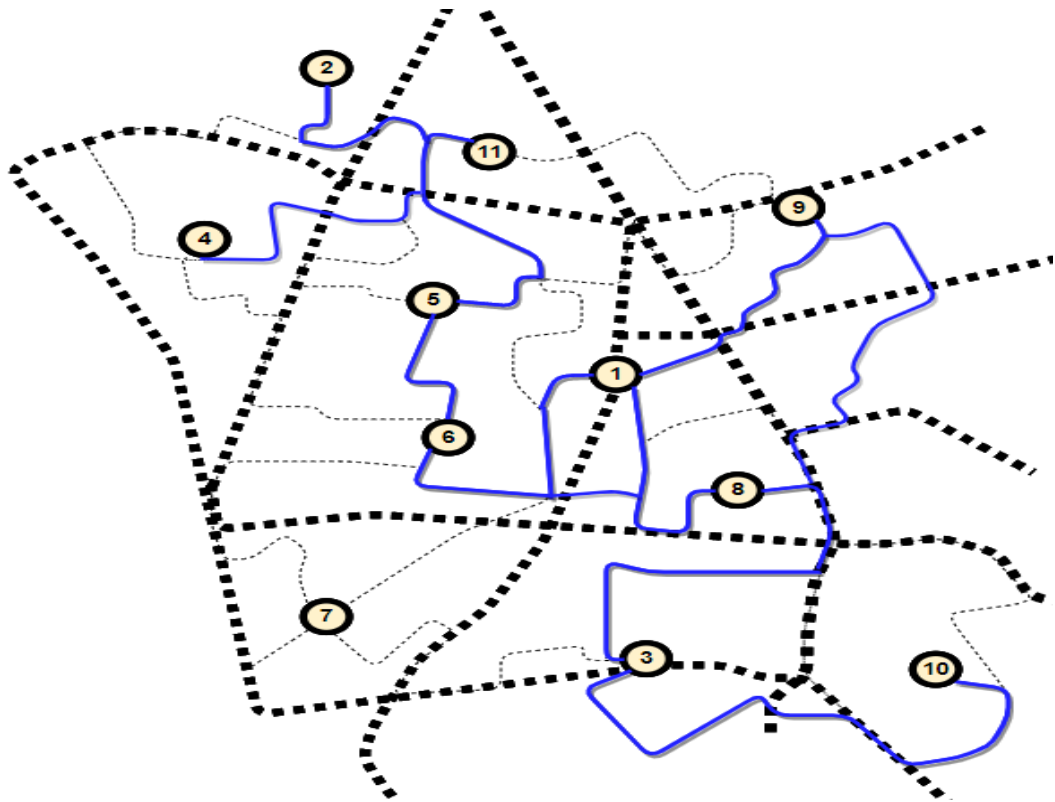

(C) At t = 20

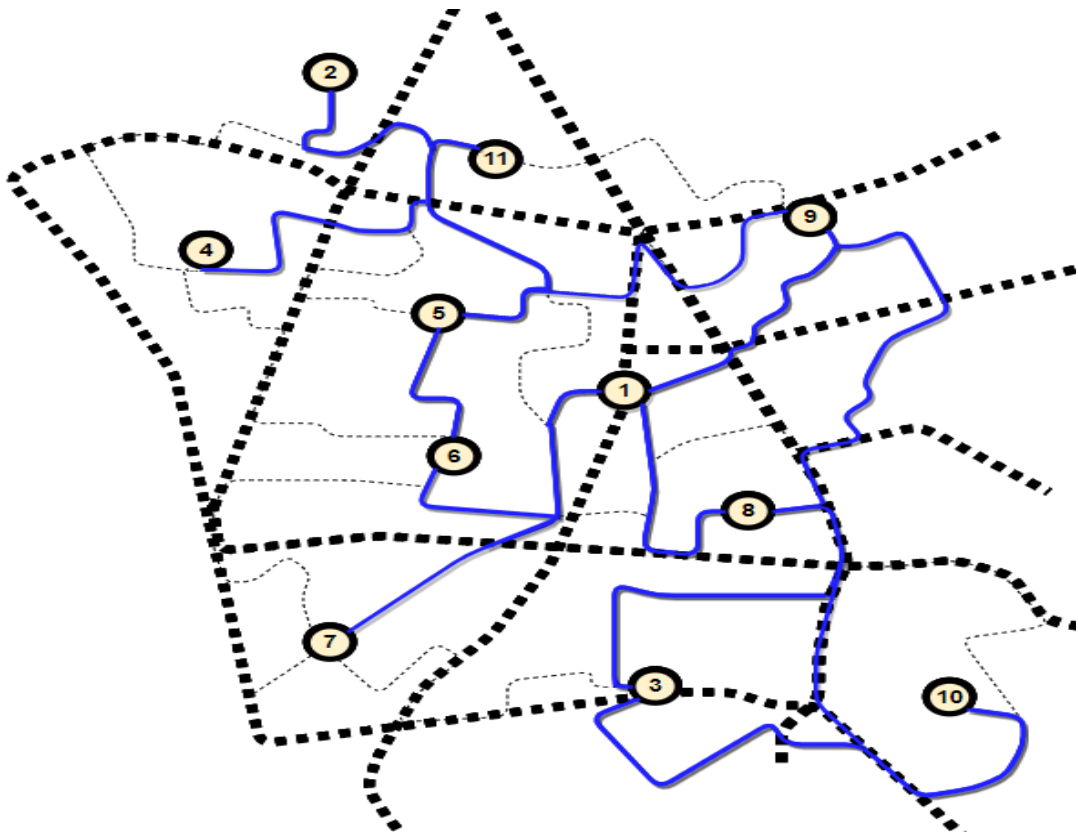

(D) At t = 50

Figure 3.4: Physarum inspired network design of 11 nodes. (A) 11 node points. The network is expanding with *t*. (C) At *t* = 20. (D) At *t* = 50.

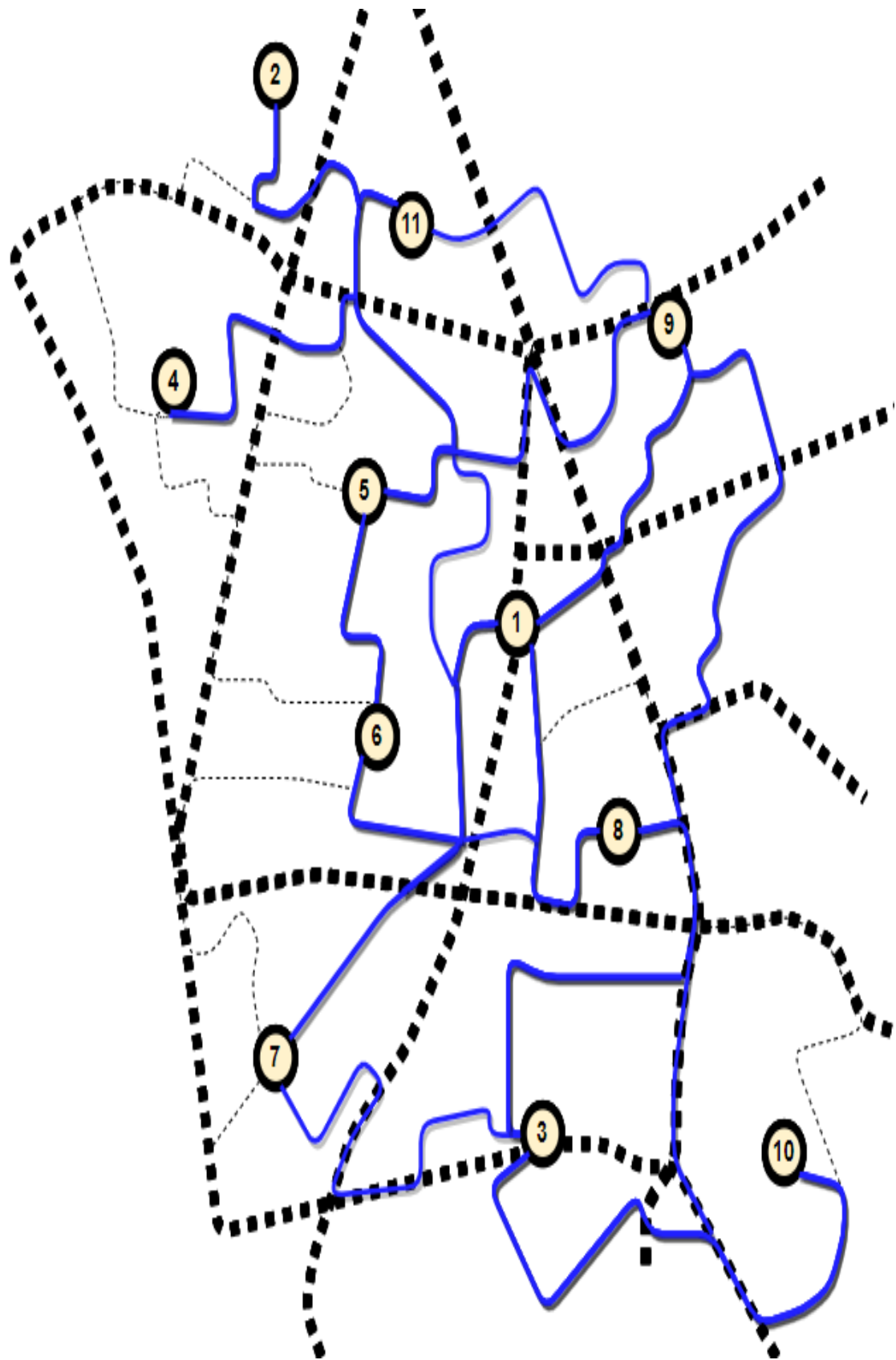

(E) Final network

Figure 3.4: Physarum inspired network design of 11 nodes. (E) Final network.

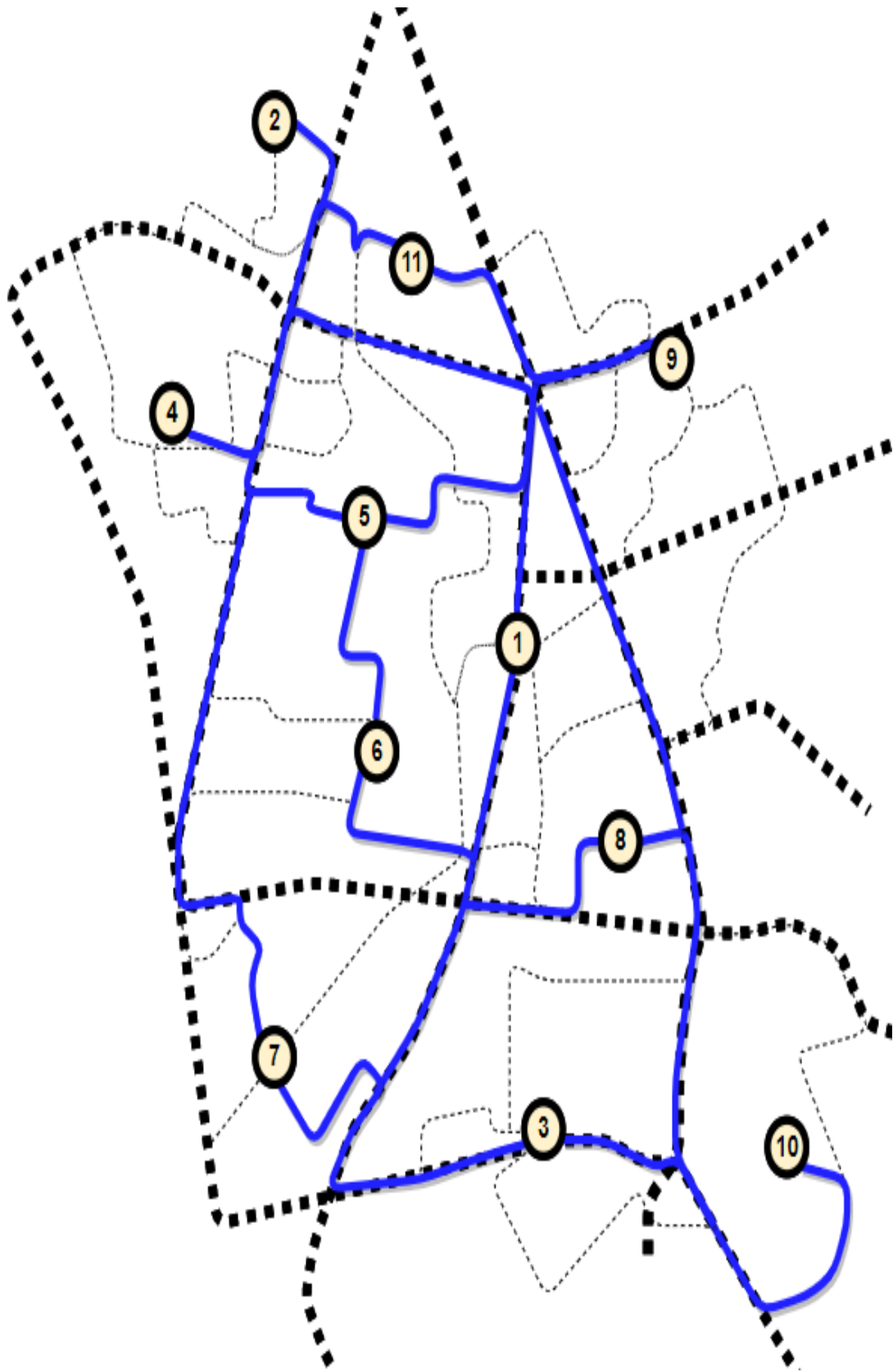

Figure 3.5: Physarum inspired network design of 11 nodes using main roads (if available).

## 3.5 Significance of Study

Modified Physarum Polycephalum Inspired Network Design Technique is used to design any real-life traffic network. In this research, the bicycle lane network is planned using this strategy in a congested mega city Dhaka. This addresses many complicated problems, including the problem of mobility.

Modified Physarum Polycephalum Inspired Network Design Technique holds a significantly different form from the existing Physarum Polycephalum Inspired Network Design Technique. Compared to previous studies it is noted that the network has only one direction between two nodes, so the stream is only flowing from one node to another, but is never flowing in the opposite direction. However, most roads have the features of double-way traffic in real traffic. There is a clear distinction between opposite directions, where flows do not interfere in two directions opposite. Apparently, in real life traffic network, the initial approach influenced by the Physarum cannot be applied. Modified Physarum Polycephalum Inspired Network Design Technique calculates flux and pressure using Eq. (3.1) and Eq. (3.4) respectively as we know that the flow of traffic is not uni-directional rather bi-directional.

# CHAPTER 4

# Experimental Studies

This chapter experimentally investigates the efficacy of the proposed modified Physarum inspired bicycle lane network design technique. For both a certain portion of Dhaka City and the entire Dhaka City, we are assuming a reduction in the number of buses, cars, taxicabs, and motorcycles. Time saving, fuel saving, user cost saving, and $CO_2$ emission reduction are calculated with some standard average measurement. In this chapter at first, we discuss the prominent points and description of this experiment, then the experimental setting which includes both parameter setting and machine description. In the third section of this chapter, we describe the experimental outcomes and explanation of achievements for both 10km ranged portion and entire Dhaka city are described.

## 4.1 Experimental Settings

In the experiment, the number of node points was 29 for both case but 77 linkages/edges are considered for prominent area of Dhaka city and 89 linkages/edges for entire Dhaka city; the length of edges($c_{ij}/L_{ij}$) which is not linear and tubular thickness($r_{ij}$) are estimated using the google map; the value of meu($\mu$) was varied from .8 to 1; for certain cases we simply ignore the tubular thickness($r_{ij}$) variations and considered a fixed conductivity value($D_{ij}$); it is assumed that, the value of pressure($P_i$) at each node point is proportional to its population in that area, and a random initial threshold is applied which is increased with the iteration.

The modified Physarum Inspired Bicycle Lane Design was implemented on Visual C++ of Visual Studio 2013. The experiments have been done on a PC (Intel Core i3-5005U CPU @ 2.00 GHz CPU, 2GB NVIDIA GeForce 940M, 4GB RAM) with Windows 10 OS.

According to Bangladesh Road Transport Authority (BRTA), on February 04, 2020, there are 127398 registered buses (including microbus, minibus), 293268 registered cars, are 724800 registered motorcycles, and 36600 registered taxicabs currently available in Dhaka city [60]. There may have some unregistered cars & buses and a lot of unregistered motorcycles & taxicabs in the city, we are not considering them. It is assumed that 40 passengers per bus, 1 person per car, taxicab and motorcycle on average. And let's assume that buses take 20km ride, cars use 10km ride, taxicabs use 100km per day and motorcycles ply 15km ride. All those rides are supposed to take place per day.

In this work, both prominent area centered with Motijheel and entire Dhaka city are considered to apply the modified Physarum inspired technique. In this section, the achievements or environmental effects of the prominent area and the entire city of Dhaka with bicycles as vehicle are respectively addressed with the proposed network.

## 4.2 Bicycle Lane Network Design in a Prominent Area

At first, a 10km selected prominent area of Dhaka city centered with Motijheel is considered to construct the network using Physarum inspired technique. In this area, we have chosen 29 vital points and numbering those from 1 to 29 arbitrarily. Here, we are considering a 10 km range because the average speed of cycling is 20kmph, so 10 km a day can be traveled easily in 30 minutes. And it can also lead to better mental health and energy by bicycling 30 minutes a day [6], [8]. Fig. 4.1 illustrates the selected portion of Dhaka city. Here, (A) depicts the portion of Dhaka city marked within the entire Dhaka city. (B) Shows the 10km range centered with Motijheel.

For 10km ranged area, it is assumed that overall 2% of users switch from car to bicycle and 10% of users switch from motorcycle to bicycle. And a 5% bus and 10% taxicab are being reduced because of using bicycle. Then the atmosphere would change significantly. This section first explains the designed networks and then calculate the time saving and then fuel and cost saving is estimated and finally $CO_2$ emission reduction is calculated.

The locations of prominent area of Dhaka city including traffic pressure is presented in Table 4.1. In this case, we assign some random values as node point's traffic pressure which is proportional to its population. For examples, Taltola, Donia, etc are less and on the other hand Motijheel, Tejgaon, etc are high traffic traffic-pressured area.

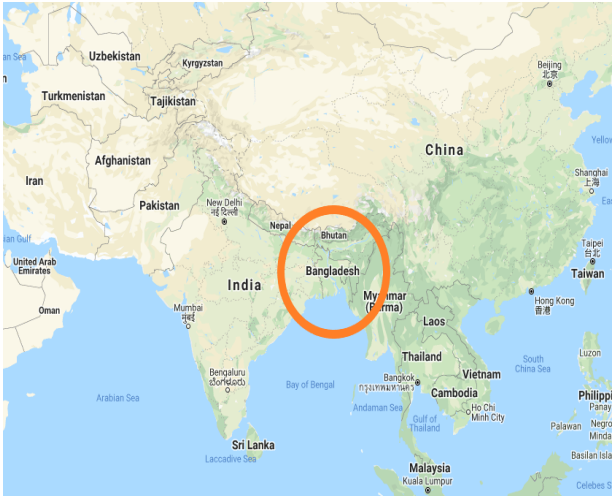

(A) Bangladesh

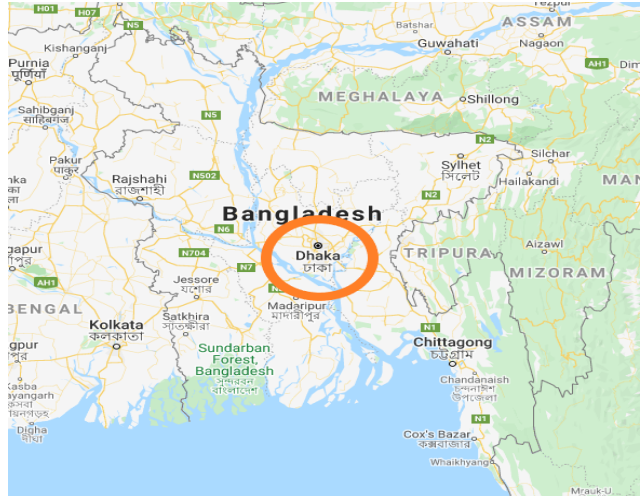

(B) Dhaka

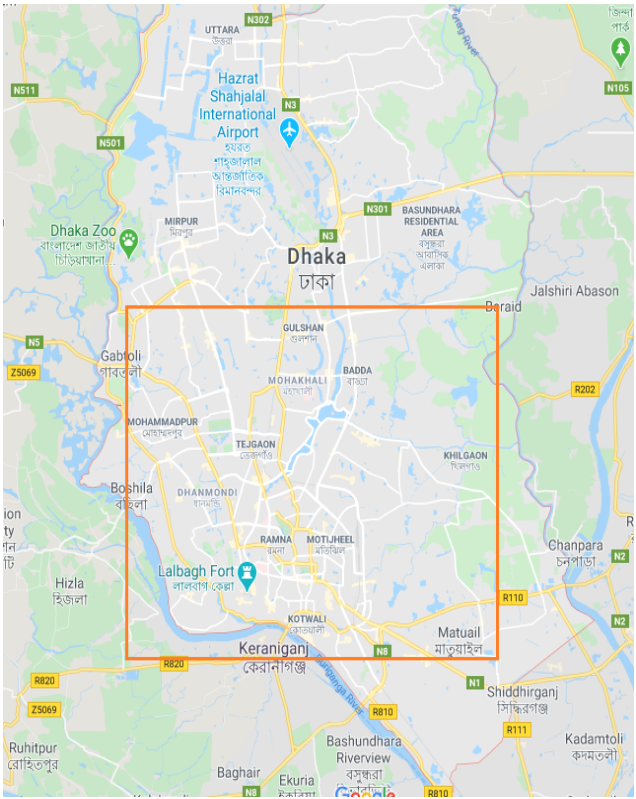

(C) Prominent area

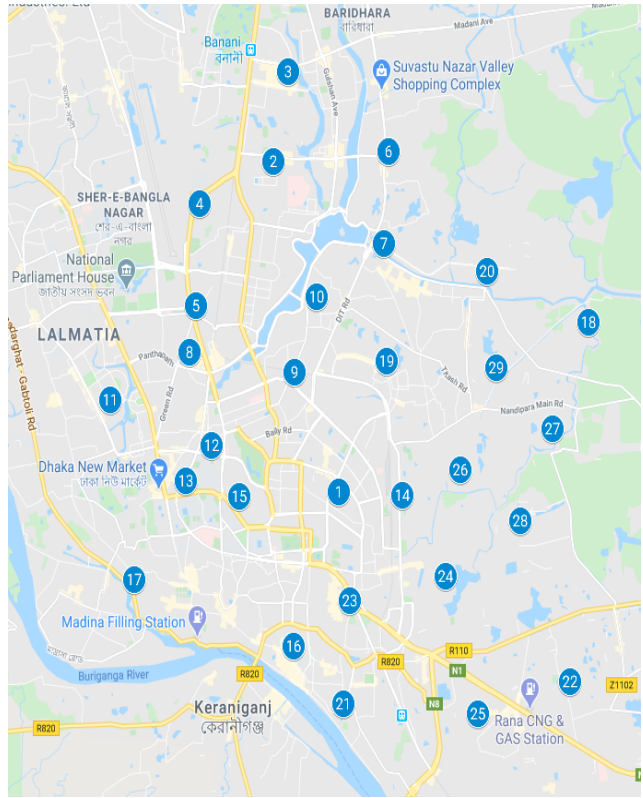

(D) Node points in Prominent area.

| | | | | | |
|---|---|---|---|---|---|
| 1 | Motijheel | 11 | Dhanmondi | 21 | Sadarghat |
| 2 | Mohakhali | 12 | Shahbag | 22 | Matuail |
| 3 | Gulshan | 13 | DU | 23 | Wari |
| 4 | Shahinbag | 14 | Kamlapur | 24 | Golapbag |
| 5 | Tejgaon | 15 | Ramna | 25 | Donia |
| 6 | Badda | 16 | Kotwali | 26 | Rajarbagh |
| 7 | EWU | 17 | Lalbagh | 27 | Sobujbagh |
| 8 | Bashundhara | 18 | Khilgaon | 28 | Green Model Town |
| 9 | Mogbazar | 19 | Taltola | 29 | Nandipara |
| 10 | Mirbag | 20 | Aftabnagar | | |

Figure 4.1: Selected Dhaka city map. (A) Bangladesh. (B) Dhaka. (C) Prominent area. (D) Node points in Prominent area.

Table 4.1: The locations of prominent area of Dhaka city including traffic pressure.

| Sl | Location | Traffic Pressure |
|----|----------|------------------|
| 01 | Motijheel | 9 |
| 02 | Mohakhali | 8 |
| 03 | Gulshan | 5 |
| 04 | Shahinbag | 8 |
| 05 | Tejgaon | 9 |
| 06 | Badda | 4 |
| 07 | EWU | 5 |
| 08 | Bashundhara | 5 |
| 09 | Mogbazar | 5 |
| 10 | Mirbag | 5 |
| 11 | Dhanmondi | 7 |
| 12 | Shahbag | 9 |
| 13 | DU | 9 |
| 14 | Kamlapur | 8 |
| 15 | Ramna | 5 |
| 16 | Kotwali | 3 |
| 17 | Lalbagh | 4 |
| 18 | Khilgaon | 6 |
| 19 | Taltola | 3 |
| 20 | Aftabnagar | 4 |
| 21 | Sadarghat | 5 |
| 22 | Matuail | 8 |
| 23 | Wari | 7 |
| 24 | Golapbag | 5 |
| 25 | Donia | 3 |
| 26 | Rajarbagh | 5 |
| 27 | Sobujbagh | 5 |
| 28 | Green Model Town | 4 |
| 29 | Nandipara | 6 |

## 4.2.1 Network Design

Here in the planned network, the distance is not the linear distance between two node points rather distance is calculated using google map. And the time is calculated considering standard speed 20kmph for bicycle in minutes.

In Fig. 4.2, (A) there are 29 node points around 10km range centered with Motijheel (node point 1) existing network with both main and local roads. Figure (B) – (E) shows that the network growing process with time (t = iteration) i.e. (B) Network in $10^{th}$ iteration; (C) Network $20^{th}$ iteration; (D) Network $50^{th}$ iteration; (E) Finale network. For those cases, we always try to avoid main roads to grow up the networks. In Fig. 4.3 Network is planned with main roads (if available).

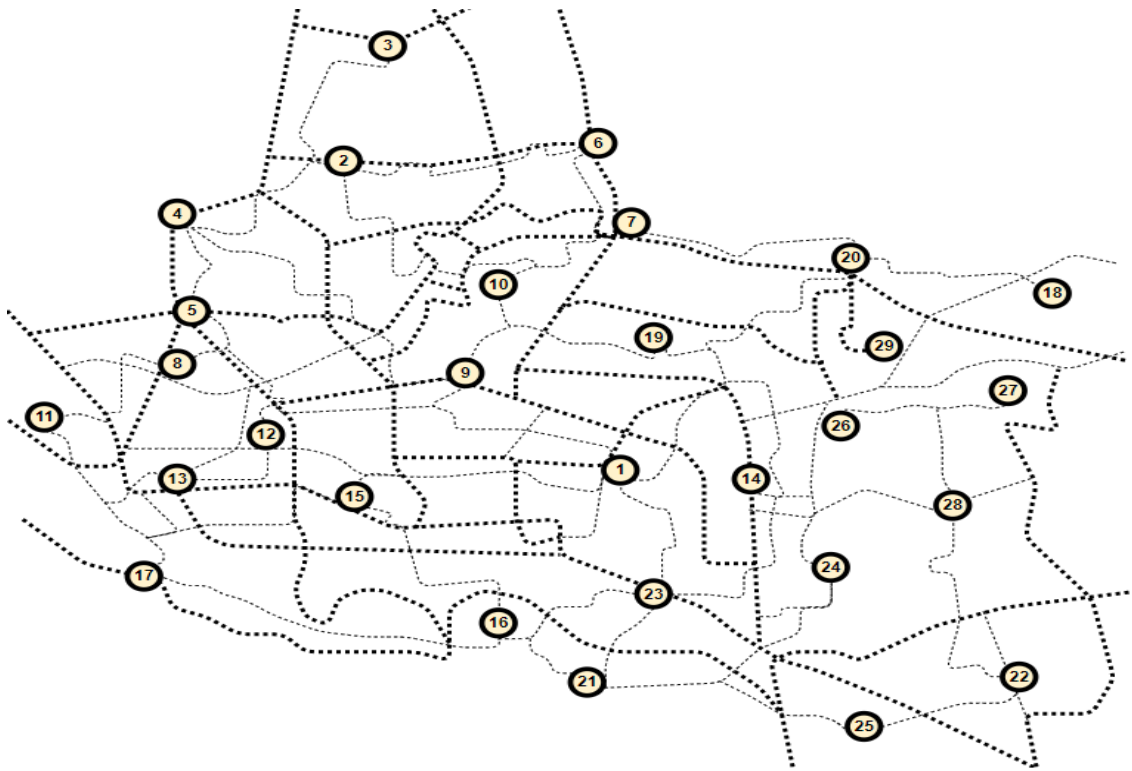

(A) 10km range existing road network

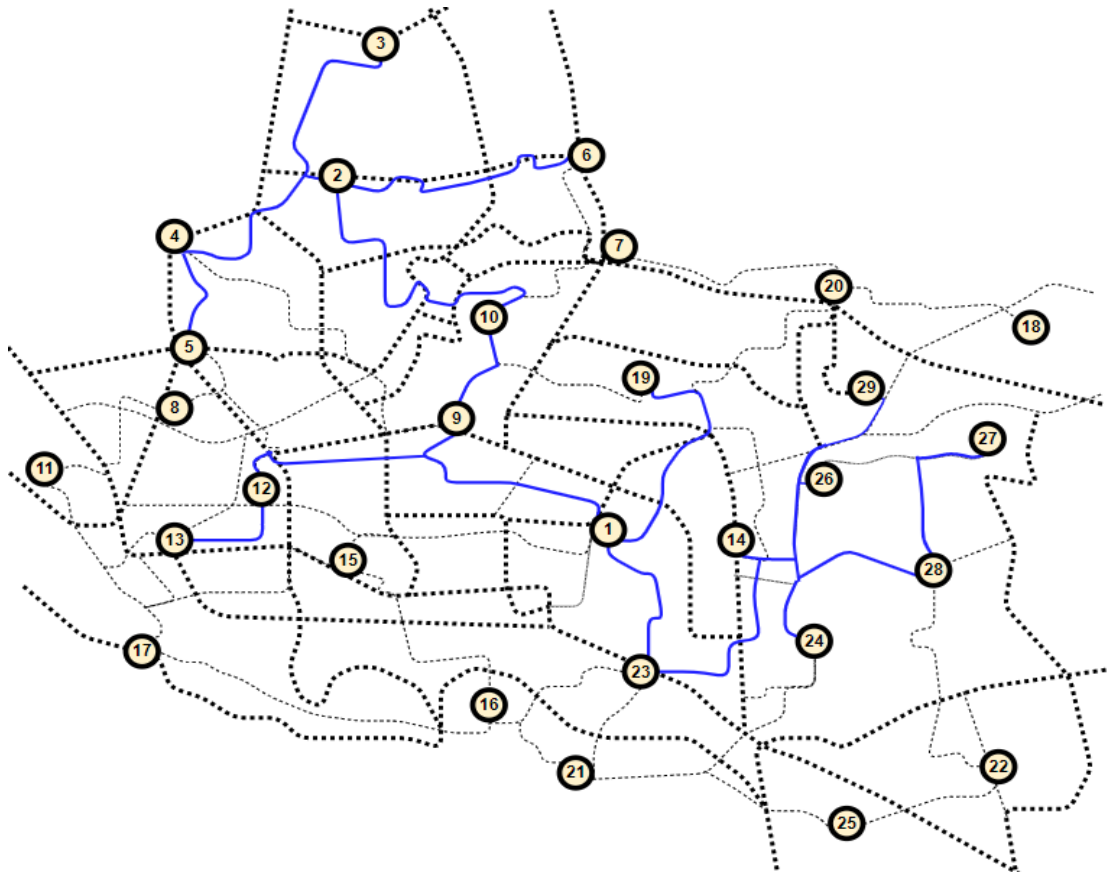

(B) When t=10

Figure 4.2: Network design using modified Physarum inspired technique. (A) 10km range existing road network. (B) When t=10.

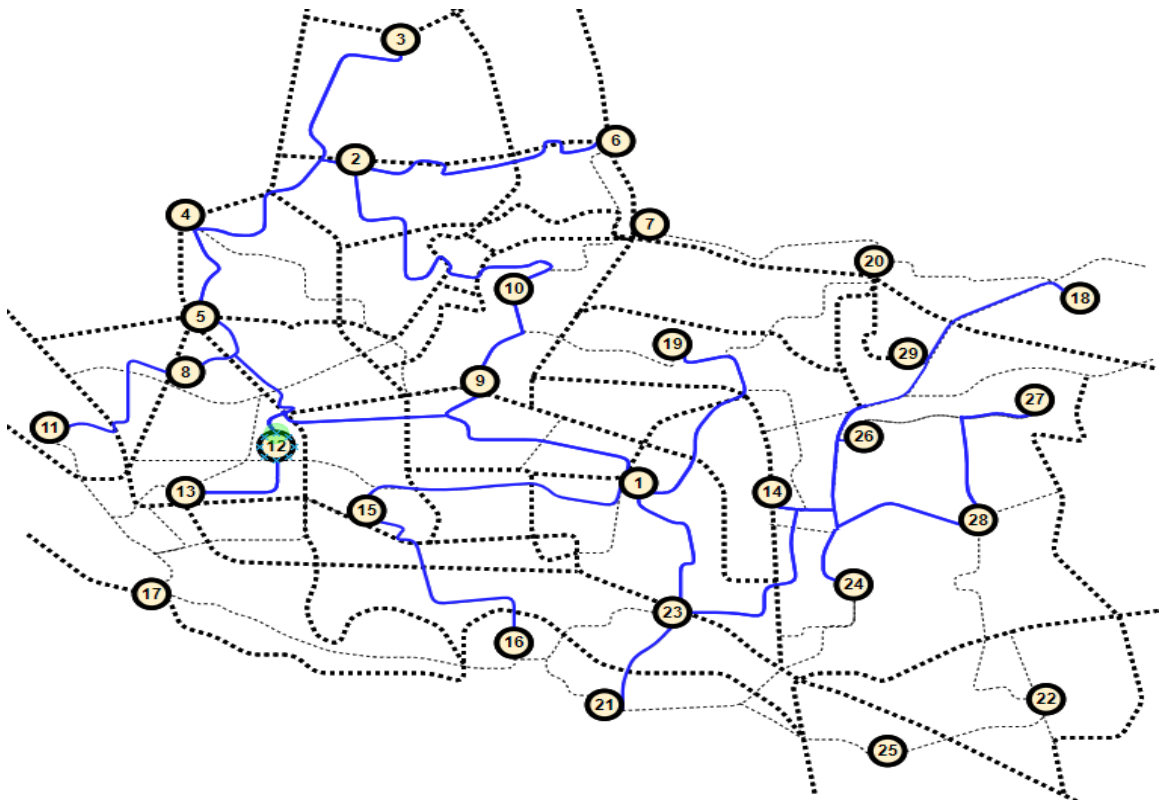

(C) When t=20

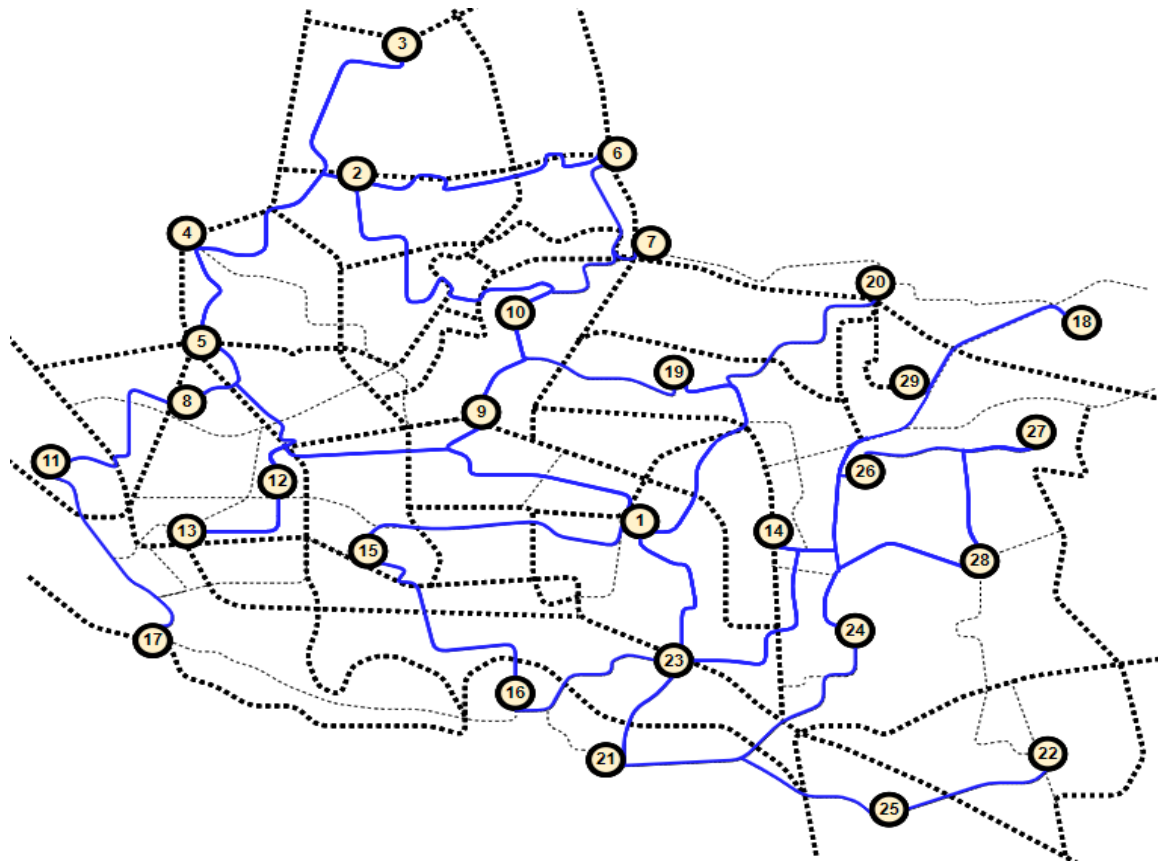

(D) When t=50

Figure 4.2: Network design using modified Physarum inspired technique. (C) When t=20. (D) When t=50.

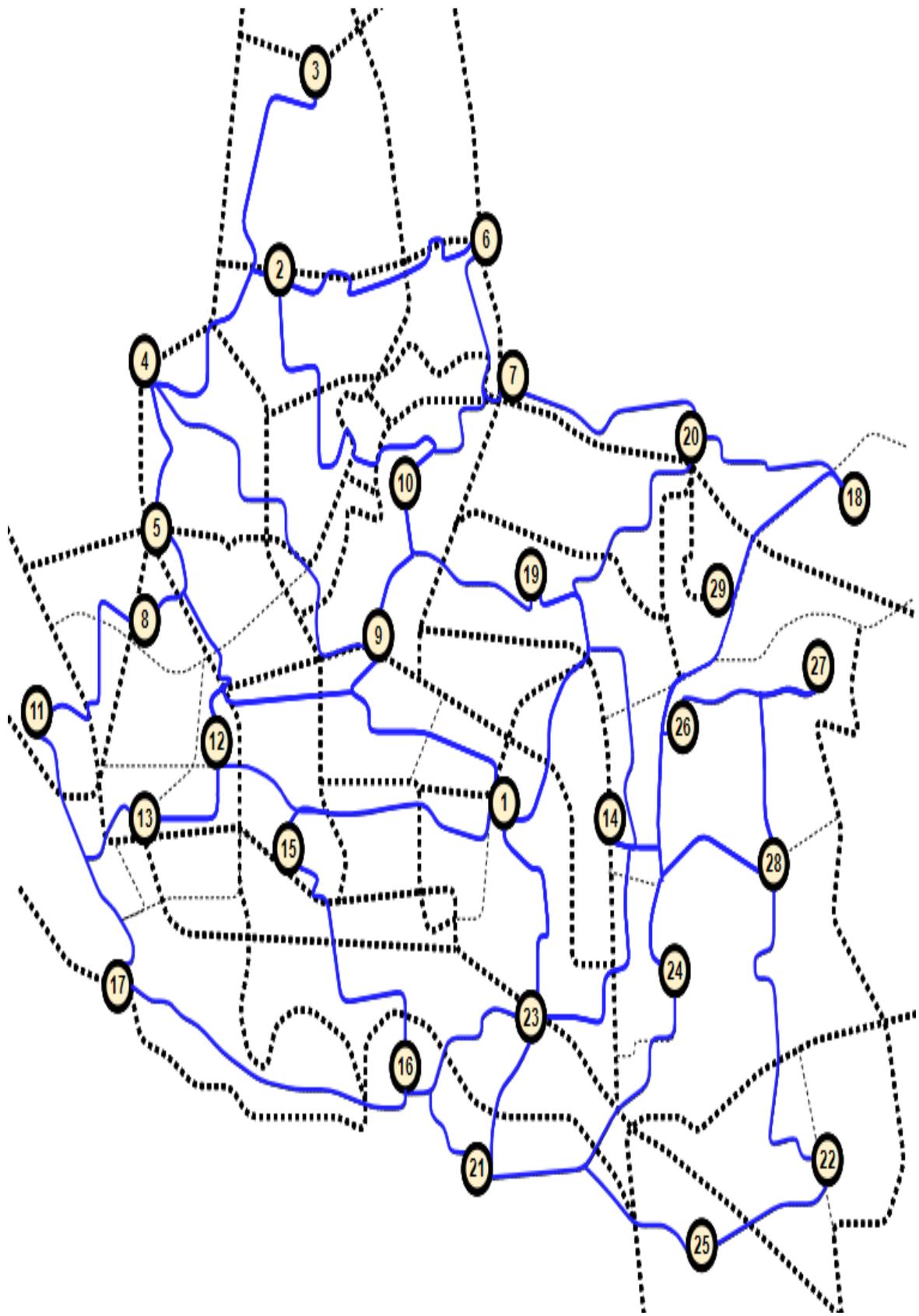

(E) Final Network

Figure 4.2: Final network design using modified Physarum inspired technique.

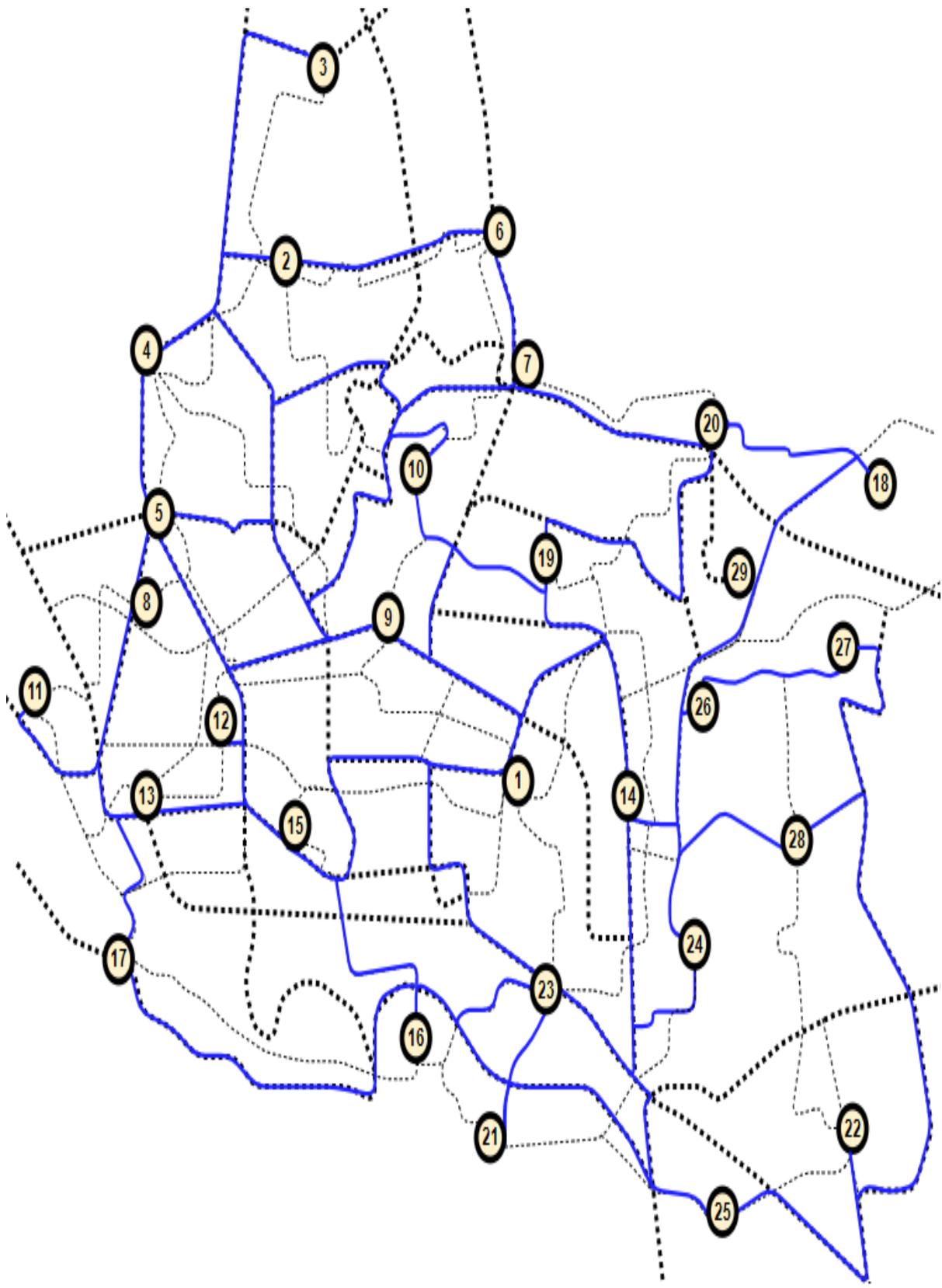

Figure 4.3: Network design using modified Physarum inspired technique using main roads (if available).

Table 4.2: Routes from node point 1.

| Des. Node Point | Routes | Distance (km) | Estimated Travel Time (min) |
|---|---|---|---|
| 2 | 1-9-10-2 | 8.55 | 25.65 |
| 3 | 1-9-10-2-3 | 11.25 | 33.75 |
| 4 | 1-9-4 | 7.3 | 21.9 |
| 5 | 1-5 | 7.5 | 22.5 |
| 6 | 1-9-10-7-6 | 8.2 | 24.6 |
| 7 | 1-9-10-7 | 6.9 | 20.7 |
| 8 | 1-8 | 6.85 | 20.55 |
| 9 | 1-9 | 2.9 | 8.7 |
| 10 | 1-9-10 | 4.4 | 13.2 |
| 11 | 1-12-13-11 | 9.6 | 28.8 |
| 12 | 1-12 | 5.3 | 15.9 |
| 13 | 1-12-13 | 6.4 | 19.2 |
| 14 | 1-14 | 5.7 | 17.1 |
| 15 | 1-15 | 2.4 | 7.2 |
| 16 | 1-23-16 | 4.5 | 13.5 |
| 17 | 1-23-16-17 | 8.1 | 24.3 |
| 18 | 1-20-18 | 9 | 27 |
| 19 | 1-19 | 3.7 | 11.1 |
| 20 | 1-20 | 6.2 | 18.6 |
| 21 | 1-23-21 | 4.5 | 13.5 |
| 22 | 1-23-21-25-22 | 10.3 | 30.9 |
| 23 | 1-23 | 2.8 | 8.4 |
| 24 | 1-14-24 | 7.15 | 21.45 |
| 25 | 1-23-21-25 | 7.1 | 21.3 |
| 26 | 1-14-26 | 6.7 | 20.1 |
| 27 | 1-14-26-27 | 9.9 | 29.7 |
| 28 | 1-14-28 | 7.9 | 23.7 |
| 29 | 1-14-26-29 | 8.1 | 24.3 |

Here the Table 4.2 shows routes of all node points from starting node point 1. For example, the path 1-9-10-2-3 means that we have to cross node points 9, 10 and 2 in order to go into destination node point 3 from starting node point 1. The minimum distance is considered for accessing any node here. For example, the destination node point 3 can be accessed using the routes 1-9-10-2-3, 1-20-7-6-2-3, 1-12-5-4-3 and so on but the minimum distanced one is considered for calculation.

### 4.2.2 Effectiveness Analysis

Driving a paddled-bicycle has a more than one beneficial environmental effect. For many cases bicycles take less time to ride, no fuel usage, saving user money, $CO_2$ emission reduction in unplanned mega city and do not contribute to air pollution, sound pollution, great relief on road traffic conditions, alleviating parking difficulties in urban areas, cause less damage to the roads, get relief of some non-communicable disease (NCDs) for have some physical exercise every day. The considering factors of this paper are only time saving, fuel and user money saving and $CO_2$ emission reduction.

### 4.2.2.1 Time Saving

As we mentioned that, according to a World Bank report, the average traffic speed in Dhaka has dropped from 21 kilometers per hour (kmph) to 7 kilometers per hour in the last 10 years, and by 2035 the speed could drop to 4 kilometers per hour, which is slower than the walking speed. In our constructed network, we mainly try to avoid main roads or minimum usage of main roads if requires. So, here traffic jam is hardly available. For convenience, we assume no traffic jam exists in the following calculation of time.

In Table 4.3, the time needed for both cars and buses are listed in three different times (at 6:00 AM, 10:00 AM, and 4:00 PM) calculated from google map and the time needed for a bicycle is always constant. Here, we measure the distance and time of all the node point from center node point 1.

The time required for a car, taxicab and motorcycle are almost the same that's why we don't mention it differently in the Table 4.3. Here we notice that at the morning 6:00 AM the travel time is less because of minimal traffic in the roads, at 10:00 AM (peak hour) when traffic jam occurs severe the need time to travel is huge and at 4:00 PM there exist traffic jam also but sometimes a bit less than peak hour. This condition is true for all types of cars, buses, taxicabs, etc. For the prominent area, the gross working hours saving are described in the following Table 4.4. So, around 152216 working hours per day can be saved using bicycle in the prominent area.

### 4.2.2.2 Fuel and Cost Saving

The cost of installation or repair is not listed here rather we considering only running cost. Because whenever we switch from car to bicycle there would be a significant reduction in costs.

Here the mileage of car and taxicab is assumed at 20kmpl and 25kmpl respectively with diesel and 65tk per liter diesel. On the other hand, cycling has no cost. The bus mileage considers 5kmpl with diesel and motorcycle has 50kmpl with petrol.

Table 4.3: Time comparison in car, bus and bicycle considering from node point 1.

| Node Point | Distance (km) | Car & Taxicab & Motor cycle | | | Bus | | | Bicycle | |
|---|---|---|---|---|---|---|---|---|---|
| | | 6:00am (min) | 10:00am (min) | 4:00pm (min) | 6:00am (min) | 10:00am (min) | 4:00pm (min) | Distance (km) | Time (min) |
| 2 | 8.6 | 18 | 39 | 36 | 20 | 43 | 40 | 8.55 | 25.65 |
| 3 | 8.9 | 19 | 42 | 39 | 21 | 46 | 43 | 11.25 | 33.75 |
| 4 | 7.1 | 14 | 33 | 30 | 16 | 36 | 33 | 7.3 | 21.9 |
| 5 | 5.1 | 12 | 27 | 26 | 14 | 30 | 29 | 7.5 | 22.5 |
| 6 | 7.6 | 15 | 36 | 36 | 17 | 39 | 39 | 8.2 | 24.6 |
| 7 | 7.6 | 15 | 36 | 36 | 17 | 39 | 39 | 6.9 | 20.7 |
| 8 | 4.8 | 11 | 24 | 24 | 13 | 27 | 27 | 6.85 | 20.55 |
| 9 | 3.6 | 9 | 18 | 15 | 11 | 20 | 17 | 2.9 | 8.7 |
| 10 | 4 | 10 | 20 | 18 | 12 | 22 | 20 | 4.4 | 13.2 |
| 11 | 6.1 | 12 | 30 | 30 | 14 | 33 | 33 | 9.6 | 28.8 |
| 12 | 3.4 | 9 | 23 | 23 | 11 | 26 | 26 | 5.3 | 15.9 |
| 13 | 3.8 | 10 | 24 | 21 | 12 | 27 | 24 | 6.4 | 19.2 |
| 14 | 1.2 | 3 | 11 | 11 | 5 | 12 | 12 | 5.7 | 17.1 |
| 15 | 3.1 | 8 | 18 | 18 | 10 | 20 | 20 | 2.4 | 7.2 |
| 16 | 3.2 | 8 | 27 | 33 | 10 | 29 | 35 | 4.5 | 13.5 |
| 17 | 6.1 | 13 | 33 | 30 | 15 | 36 | 33 | 8.1 | 24.3 |
| 18 | 7.2 | 14 | 53 | 48 | 16 | 56 | 51 | 9 | 27 |
| 19 | 3 | 8 | 21 | 18 | 10 | 23 | 20 | 3.7 | 11.1 |
| 20 | 10 | 24 | 42 | 38 | 26 | 47 | 43 | 6.2 | 18.6 |
| 21 | 3.5 | 9 | 36 | 36 | 11 | 38 | 38 | 4.5 | 13.5 |
| 22 | 8.3 | 18 | 33 | 30 | 20 | 37 | 34 | 10.3 | 30.9 |
| 23 | 2.5 | 6 | 21 | 21 | 8 | 23 | 23 | 2.8 | 8.4 |
| 24 | 4.2 | 10 | 27 | 27 | 12 | 30 | 30 | 7.15 | 21.45 |
| 25 | 5.7 | 13 | 29 | 27 | 15 | 32 | 30 | 7.1 | 21.3 |
| 26 | 4.7 | 11 | 29 | 29 | 13 | 32 | 32 | 6.7 | 20.1 |
| 27 | 3.1 | 8 | 15 | 12 | 10 | 17 | 14 | 9.9 | 29.7 |
| 28 | 5.8 | 12 | 42 | 39 | 14 | 45 | 42 | 7.9 | 23.7 |
| 29 | 5.4 | 12 | 33 | 33 | 14 | 36 | 36 | 8.1 | 24.3 |

Table 4.4: Time saving per day.

| Transit | % of Transit Reduction | Transit Reduces | Riding Distance (km) | Time saves (min) |
|---|---|---|---|---|
| Bus | 5% | 6370 | 127400 | 709800 |
| Car | 2% | 5865 | 58650 | 326764 |
| Taxicab | 10% | 3660 | 366000 | 2039143 |
| Motor cycle | 10% | 72480 | 1087200 | 6057257 |
| | | | Total time saving | 9132964 |

In Table 4.4, the needed fuel for cars, taxicabs, motorcycles, and buses is calculated and there is no running cost & fuel cost for the bicycle. In case of bus, the per km fare is fixed by BRTA of 1.7tk and we assumed that taxicab fare is 50tk per km. Here, the distance of all the node points are measured from center node point 1.

Table 4.5: Cost comparison in car, bus and bicycle considering from node point 1.

| Node Point | Distance (km) | Car | | Taxicabs | | Motor cycle | | Bus | | Bicycle | |
|---|---|---|---|---|---|---|---|---|---|---|---|
| | | Fuel (litre) | User Cost (tk) | Fuel (litre) | User Cost (tk) | Fuel (litre) | User Cost (tk) | Fuel (litre) | User Cost (tk) | Distance (km) | Cost (tk) |
| 2 | 8.6 | 0.43 | 27.95 | 0.34 | 430 | 0.17 | 15.31 | | 14.62 | 8.55 | |
| 3 | 8.9 | 0.45 | 28.93 | 0.36 | 445 | 0.18 | 15.84 | | 15.13 | 11.25 | |
| 4 | 7.1 | 0.36 | 23.08 | 0.28 | 355 | 0.14 | 12.64 | | 12.07 | 7.3 | |
| 5 | 5.1 | 0.26 | 16.58 | 0.2 | 255 | 0.1 | 9.08 | As we assume bicycles reduces 5% of the total bus in Dhaka city. The fuel consumption is reduced = 6370×20 ×1/5 litres = 25480 litres | 8.67 | 7.5 | |
| 6 | 6.8 | 0.34 | 22.1 | 0.27 | 340 | 0.14 | 12.1 | | 11.56 | 8.2 | |
| 7 | 7.6 | 0.38 | 24.7 | 0.3 | 380 | 0.15 | 13.53 | | 12.92 | 6.9 | |
| 8 | 4.8 | 0.24 | 15.6 | 0.19 | 240 | 0.1 | 8.54 | | 8.16 | 6.85 | |
| 9 | 3.6 | 0.18 | 11.7 | 0.14 | 180 | 0.07 | 6.41 | | 6.12 | 2.9 | |
| 10 | 4 | 0.2 | 13 | 0.16 | 200 | 0.08 | 7.12 | | 6.8 | 4.4 | |
| 11 | 6.1 | 0.31 | 19.83 | 0.24 | 305 | 0.12 | 10.86 | | 10.37 | 9.6 | |
| 12 | 3.4 | 0.17 | 11.05 | 0.14 | 170 | 0.07 | 6.05 | | 5.78 | 5.3 | |
| 13 | 3.8 | 0.19 | 12.35 | 0.15 | 190 | 0.08 | 6.76 | | 6.46 | 6.4 | |
| 14 | 1.2 | 0.06 | 3.9 | 0.05 | 60 | 0.02 | 2.14 | | 2.04 | 5.7 | N/A |
| 15 | 3.1 | 0.16 | 10.08 | 0.12 | 155 | 0.06 | 5.52 | | 5.27 | 2.4 | |
| 16 | 3.2 | 0.16 | 10.4 | 0.13 | 160 | 0.06 | 5.7 | | 5.44 | 4.5 | |
| 17 | 6.1 | 0.31 | 19.83 | 0.24 | 305 | 0.12 | 10.86 | | 10.37 | 8.1 | |
| 18 | 7.2 | 0.36 | 23.4 | 0.29 | 360 | 0.14 | 12.82 | | 12.24 | 9 | |
| 19 | 3 | 0.15 | 9.75 | 0.12 | 150 | 0.06 | 5.34 | | 5.1 | 3.7 | |
| 20 | 10 | 0.5 | 32.5 | 0.4 | 500 | 0.2 | 17.8 | | 17 | 6.2 | |
| 21 | 3.5 | 0.18 | 11.38 | 0.14 | 175 | 0.07 | 6.23 | | 5.95 | 4.5 | |
| 22 | 8.3 | 0.42 | 26.98 | 0.33 | 415 | 0.17 | 14.77 | | 14.11 | 10.3 | |
| 23 | 2.5 | 0.13 | 8.13 | 0.1 | 125 | 0.05 | 4.45 | | 4.25 | 2.8 | |
| 24 | 4.2 | 0.21 | 13.65 | 0.17 | 210 | 0.08 | 7.48 | | 7.14 | 7.15 | |
| 25 | 5.7 | 0.29 | 18.53 | 0.23 | 285 | 0.11 | 10.15 | | 9.69 | 7.1 | |
| 26 | 4.7 | 0.24 | 15.28 | 0.19 | 235 | 0.09 | 8.37 | | 7.99 | 6.7 | |
| 27 | 3.1 | 0.16 | 10.08 | 0.12 | 155 | 0.06 | 5.52 | | 5.27 | 9.9 | |
| 28 | 5.8 | 0.29 | 18.85 | 0.23 | 290 | 0.12 | 10.32 | | 9.86 | 7.9 | |
| 29 | 5.4 | 0.27 | 17.55 | 0.22 | 270 | 0.11 | 9.61 | | 9.18 | 8.1 | |

The gross fuel saving and user cost saving for motijheel area are calculated for buses, cars, taxicabs, and motorcycles and described in the Table 4.6. So, around 64797 liters fuel and 20.6 million user money costs can be saved per day using bicycle in the motijheel area.

Table 4.6: Cost saving per day.

| Transit | % of Transit Reduction | Transit Reduces | Riding Distance (km) | Fuel (litres) | User Cost (tk) |
|---|---|---|---|---|---|
| Bus | 5% | 6370 | 127400 | 25480 | 216580 |
| Car | 2% | 5865 | 58650 | 2933 | 190613 |
| Taxicab | 10% | 3660 | 366000 | 14640 | 18300000 |
| Motor cycle | 10% | 72480 | 1087200 | 21744 | 1935216 |
| | | | Total cost saving | 64797 | 20642409 |

### 4.2.2.3 $CO_2$ Emission Reduction

In the calculation, 887 g/km, 258 g/km, 237 g/km and 40 g/km are the considered amount of $CO_2$ emission in 1km ride of bus, car, taxicab and motorcycle respectively. For motijheel, the gross $CO_2$ emission reduction is described in the following Table 4.7.

Table 4.7: $CO_2$ emission reduction per day.

| Transit | % of Transit Reduction | Transit Reduces | Riding Distance (km) | $CO_2$ Emission (g) |
|---|---|---|---|---|
| Bus | 5% | 6370 | 127400 | $1.13 \times 10^8$ |
| Car | 2% | 5865 | 58650 | $1.51 \times 10^7$ |
| Taxicab | 10% | 3660 | 366000 | $8.67 \times 10^7$ |
| Motor cycle | 10% | 72480 | 1087200 | $4.35 \times 10^7$ |
| | | | Total $CO_2$ saving | $2.58 \times 10^8$ |

In total, around $6.5 \times 10^5$ kg $CO_2$ emission is reduced in the motijheel area per day.

## 4.3 Bicycle Lane Network Design for Entire Dhaka City

Here, 29 important locations of entire Dhaka city are selected and numbering those from 1 to 29 randomly, which are depicted in following Fig. 4.4. As it is considered that the paddled-bicycle for routing 10km ranged area, it is not feasible to move through all over the Dhaka city using paddled-bicycle. But in case of electric bicycle, it is possible. That's why electric bicycles are considered as vehicles for entire Dhaka city.

The general data of Dhaka city including location, population and traffic pressure are described in the Table 4.8. In this case, the consideration is that the node point's traffic pressure is proportional to its population.

For entire Dhaka city, it is assumed that 5% of users switch from car to electrical motorcycle or bicycle and 50% of users switch from motorcycle to electrical motorcycle or bicycle. And a 10% bus and 20% taxicab are being reduced because of using electric motorcycles or bicycles. Then there will be some noteworthy change in the environment.

### 4.3.1 Network Design

The constructed network of selected 29 points of Dhaka city is illustrated in Fig. 4.5. In the Table 4.9, routes are shown to all node points from starting node point 7. For example, the route 7-6-24-5-1 means that we have to pass node points 6, 26 and 5 in order to arrive destination node point 1 from starting node point 7. We are considering the minimum distance for accessing any node here. For example, the destination node point 1 can be accessed using the routes 7-6-24-5-1, 7-6-4-26-5-1, 7-21-22-2-24-5-1 and so on. But the minimum distanced route is considered for calculation.

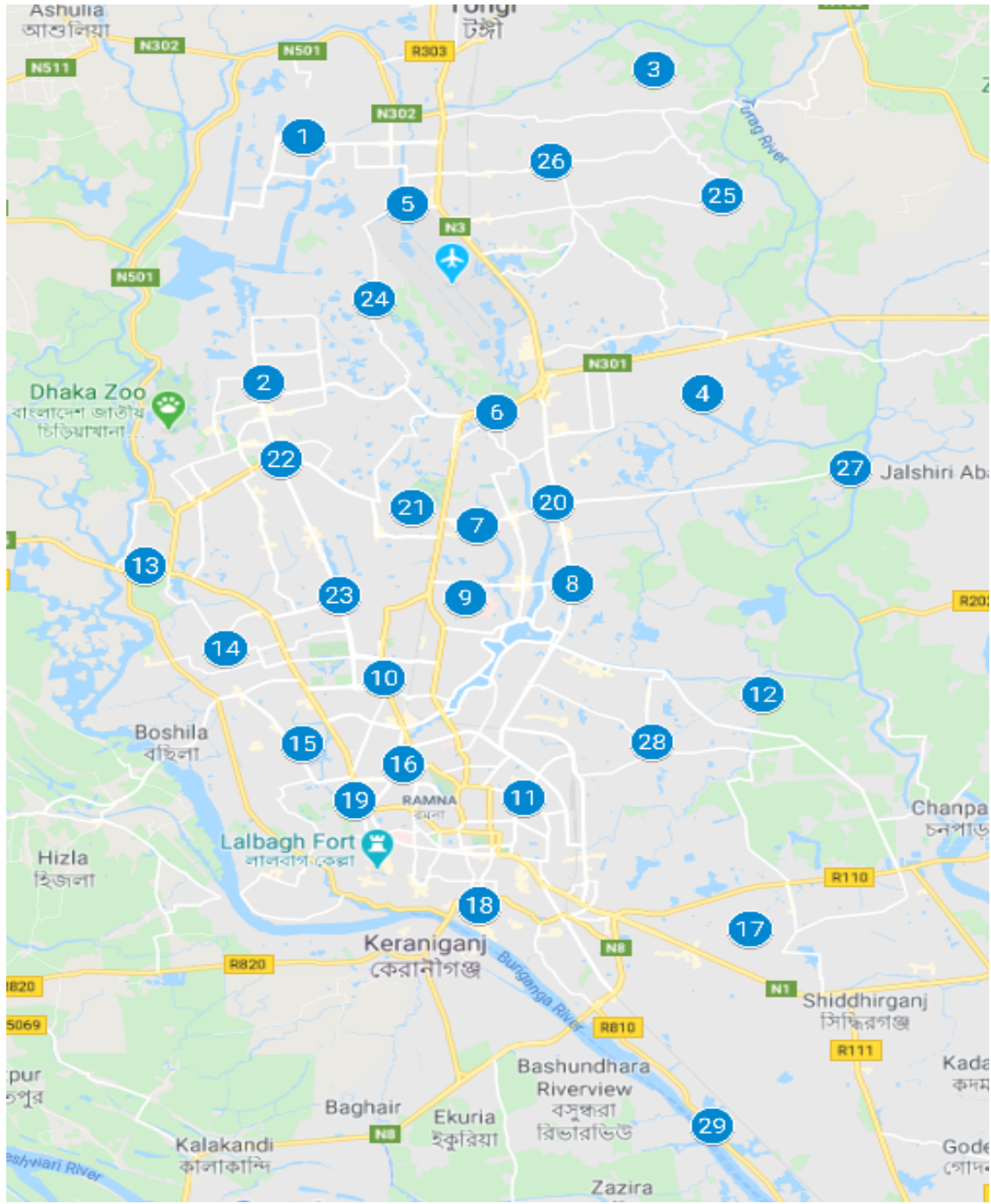

| | | |
|---|---|---|
| 01 Uttara | 11 Motijheel | 21 Banani |
| 02 Mirpur | 12 Khilgoan | 22 Monipur |
| 03 Uttara ABM City | 13 Gabtoli | 23 Sher-E-Bangla Nagar |
| 04 Basundhara R/A | 14 Mohammadpur | 24 Airport |
| 05 Khilkhet | 15 Dhanmondi | 25 Poradia |
| 06 Cantonment | 16 Shahbagh | 26 Madarbari |
| 07 Gulshan | 17 Matuail | 27 Beraid |
| 08 Badda | 18 Kotwali | 28 NutanPara |
| 09 Mohakhali | 19 New Market | 29 Pagla Rail station |
| 10 Tejgaon | 20 Baridhara | |

Figure 4.3: Entire Dhaka city.

Table 4.8: The general data on locations in Dhaka city, including population, area, and traffic pressure.

| Sl | Location | Population | Area (km²) | Traffic Pressure |
|---|---|---|---|---|
| 01 | Uttara | 345,097 | 36.91 | 10 |
| 02 | Mirpur | 274,530 | 4.71 | 8 |
| 03 | Uttara ABM City | 145,097 | 27.91 | 5 |
| 04 | Bashundhara R/A | 274,200 | 13.54 | 8 |
| 05 | Khilkhet | 130,053 | 15.88 | 5 |
| 06 | Cantonment | 117,464 | 14.47 | 4 |
| 07 | Gulshan | 145,969 | 8.85 | 5 |
| 08 | Badda | 157,924 | 16.78 | 5 |
| 09 | Mohakhali | 145,969 | 8.85 | 5 |
| 10 | Tejgaon | 148,255 | 2.46 | 5 |
| 11 | Motijheel | 225,999 | 3.69 | 7 |
| 12 | Khilgaon | 327,717 | 13.8 | 9 |
| 13 | Gabtoli | 198,723 | 4.98 | 6 |
| 14 | Mohammadpur | 355,843 | 11.65 | 10 |
| 15 | Dhanmondi | 147,643 | 2.86 | 5 |
| 16 | Shahbag | 74,113 | 3.49 | 3 |
| 17 | Matuail | 125,312 | 19.36 | 4 |
| 18 | Kotwali | 210,504 | 0.67 | 6 |
| 19 | New Market | 66,439 | 1.67 | 2 |
| 20 | Baridhara | 105,969 | 5.45 | 4 |
| 21 | Banani | 145,969 | 8.85 | 5 |
| 22 | Monipur | 274,530 | 4.71 | 8 |
| 23 | Sher-E-Bangla Nagar | 248,871 | 5.25 | 7 |
| 24 | Airport | 130,053 | 15.88 | 5 |
| 25 | Poradia | 52,014 | 20.09 | 2 |
| 26 | Madarbari | 93,153 | 12.65 | 3 |
| 27 | Beraid | 157,924 | 16.78 | 5 |
| 28 | NutanPara | 125,312 | 19.36 | 4 |
| 29 | Pagla Rail station | 194,019 | 246.21 | 6 |

### 4.3.2 Effectiveness Analysis

Using an electric-bicycle has a more than one beneficial environmental effect. For most of the cases bicycles take less time to ride, no fuel usage, saving user money, $CO_2$ emission reduction in unplanned mega city and very less contribution to air pollution, sound pollution, great relief on road traffic conditions, alleviating parking difficulties in urban areas, cause less damage to the roads, get relief of some non-communicable disease (NCDs) for have some physical exercise every day. The considering factors of this paper is only time saving, fuel and user money saving and CO2 emission reduction.

Table 4.9: Routes from node point 7.

| Des. Node Point | Routes | Distance (km) | Estimated Travel Time(min) |
|---|---|---|---|
| 1 | 7-6-24-5-1 | 15.65 | 23.48 |
| 2 | 7-21-22-2 | 9.89 | 14.84 |
| 3 | 7-6-24-5-26-3 | 19.51 | 29.27 |
| 4 | 7-6-4 | 9.02 | 13.53 |
| 5 | 7-6-24-5 | 11.9 | 17.85 |
| 6 | 7-6 | 3.95 | 5.93 |
| 8 | 7-20-8 | 3.17 | 4.76 |
| 9 | 7-9 | 2.17 | 3.26 |
| 10 | 7-9-10 | 5.98 | 8.97 |
| 11 | 7-9-10-16-11 | 10.61 | 15.92 |
| 12 | 7-20-12 | 9.04 | 13.56 |
| 13 | 7-21-13 | 9.44 | 14.16 |
| 14 | 7-23-14 | 7.31 | 10.97 |
| 15 | 7-23-14-15 | 11.55 | 17.33 |
| 16 | 7-9-10-16 | 8.7 | 13.05 |
| 17 | 7-9-28-17 | 15.61 | 23.42 |
| 18 | 1-20-18 | 14.78 | 22.17 |
| 19 | 7-9-10-19 | 9.87 | 14.81 |
| 20 | 7-20 | 2.12 | 3.18 |
| 21 | 7-21 | 1.76 | 2.64 |
| 22 | 7-21-22 | 7.34 | 11.01 |
| 23 | 7-23 | 4.07 | 6.11 |
| 24 | 7-6-24 | 8.55 | 12.83 |
| 25 | 7-6-4-25 | 18.72 | 28.08 |
| 26 | 7-6-24-5-26 | 15.5 | 23.25 |
| 27 | 7-20-27 | 11.2 | 16.8 |
| 28 | 7-9-28 | 9.41 | 14.12 |
| 29 | 7-9-28-17-29 | 22.02 | 33.03 |

## 4.3.2.1 Time Saving

In the following measurements, 7 kilometers per hour (kmph) for the buses, cars, taxicabs & motorcycles which is the average speed of traffic in Dhaka city and 40 km/h for electric bicycles using our constructed network. We are taken into consideration our constructed network as jam-free. In the Table 4.10, the time necessary for cars and buses are listed in three different times (at 6:00 AM, 10:00 AM, and 4:00 PM) calculated from google map and the time needed for a bicycle is always constant. The required time for taxicabs and motorcycles is considered to be the same as the time needed for a car. Here, we measure the distance and time of all the node point from center node point 7.

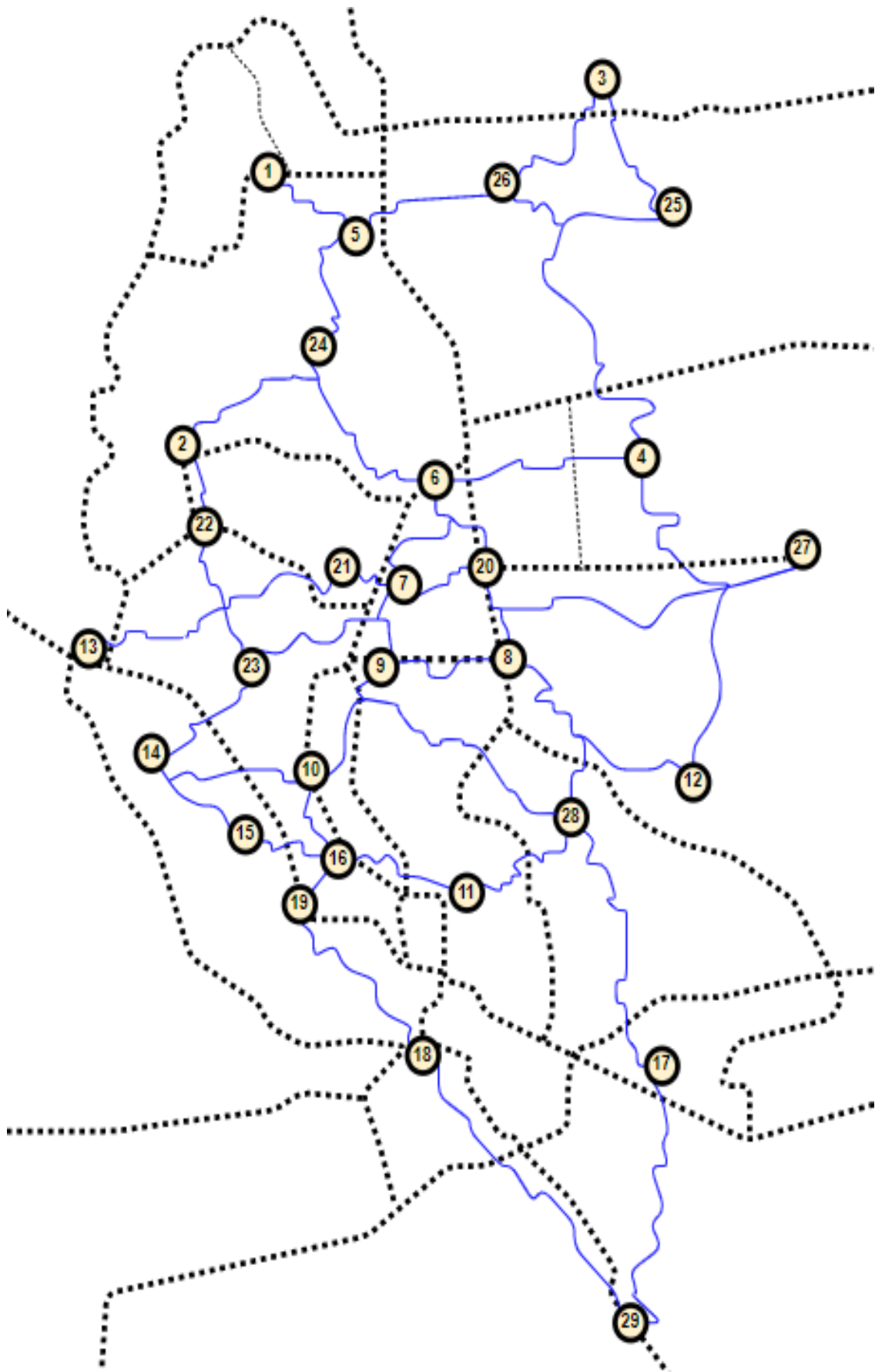

Figure 4.4: Constructed network 29 points of Dhaka city.

Table 4.10: Time comparison in car, bus and bicycle considering from node point 7.

| Node Point | Distance (km) | Car & Taxicab & Motor cycle | | | Bus | | | Bicycle | |
|---|---|---|---|---|---|---|---|---|---|
| | | 6:00am (min) | 10:00am (min) | 4:00pm (min) | 6:00am (min) | 10:00am (min) | 4:00pm (min) | Distance (km) | Time (min) |
| 1 | 14 | 24 | 42 | 45 | 26 | 50 | 51 | 15.65 | 23.48 |
| 2 | 7.5 | 16 | 33 | 36 | 18 | 43 | 42 | 9.89 | 14.84 |
| 3 | 18.2 | 40 | 75 | 75 | 42 | 82 | 81 | 19.51 | 29.27 |
| 4 | 10.2 | 22 | 42 | 36 | 22 | 42 | 42 | 9.02 | 13.53 |
| 5 | 8.7 | 12 | 24 | 24 | 14 | 31 | 30 | 11.9 | 17.85 |
| 6 | 6 | 10 | 26 | 24 | 12 | 32 | 30 | 3.95 | 5.93 |
| 8 | 4.7 | 8 | 30 | 27 | 9 | 35 | 32 | 3.17 | 4.76 |
| 9 | 3.5 | 7 | 24 | 22 | 9 | 28 | 24 | 2.17 | 3.26 |
| 10 | 5.5 | 10 | 25 | 24 | 12 | 31 | 28 | 5.98 | 8.97 |
| 11 | 9.1 | 18 | 45 | 42 | 20 | 52 | 48 | 10.61 | 15.92 |
| 12 | 12 | 40 | 60 | 57 | 42 | 68 | 65 | 9.04 | 13.56 |
| 13 | 9.7 | 22 | 44 | 45 | 24 | 51 | 51 | 9.44 | 14.16 |
| 14 | 8.3 | 14 | 36 | 33 | 16 | 44 | 39 | 7.31 | 10.97 |
| 15 | 8.7 | 16 | 33 | 36 | 17 | 41 | 42 | 11.55 | 17.33 |
| 16 | 8 | 16 | 39 | 39 | 18 | 47 | 45 | 8.7 | 13.05 |
| 17 | 16.6 | 30 | 63 | 60 | 32 | 71 | 66 | 15.61 | 23.42 |
| 18 | 11.4 | 24 | 62 | 60 | 26 | 60 | 66 | 14.78 | 22.17 |
| 19 | 8.7 | 18 | 42 | 42 | 20 | 50 | 48 | 9.87 | 14.81 |
| 20 | 2.2 | 5 | 12 | 11 | 6 | 15 | 14 | 2.12 | 3.18 |
| 21 | 2.3 | 5 | 17 | 15 | 7 | 17 | 15 | 1.76 | 2.64 |
| 22 | 6.8 | 20 | 36 | 33 | 22 | 44 | 39 | 7.34 | 11.01 |
| 23 | 6.6 | 12 | 24 | 23 | 14 | 32 | 29 | 4.07 | 6.11 |
| 24 | 8.7 | 12 | 24 | 24 | 14 | 32 | 30 | 8.55 | 12.83 |
| 25 | 15.9 | 30 | 68 | 71 | 32 | 75 | 77 | 18.72 | 28.08 |
| 26 | 14.7 | 30 | 63 | 60 | 32 | 71 | 66 | 15.5 | 23.25 |
| 27 | 8.5 | 14 | 36 | 33 | 16 | 45 | 39 | 11.2 | 16.8 |
| 28 | 10.2 | 24 | 63 | 60 | 26 | 71 | 66 | 9.41 | 14.12 |
| 29 | 19.1 | 35 | 79.5 | 75 | 37 | 87 | 81 | 22.02 | 33.03 |

Table 4.11: Time saving per day.

| Transit | % of Transit Reduction | Transit Reduces | Riding Distance (km) | Time Saves (min) |
|---|---|---|---|---|
| Bus | 10% | 12740 | 254796 | $7.2 \times 10^7$ |
| Car | 5% | 14663 | 146630 | $1.04 \times 10^6$ |
| Taxicab | 20% | 7320 | 732000 | $5.2 \times 10^6$ |
| Motor cycle | 50% | 362400 | 5436000 | $3.8 \times 10^7$ |
| | | | Total time saving | $1.16 \times 10^8$ |

In the Table 4.10, we have noticed that at the morning 6:00 AM the travel time is less because of minimal traffic in the roads, at 10:00 AM (peak hour) when traffic jam occurs severe amount the need time to travel is huge and at 4:00 PM there exist traffic jam also but sometimes a bit less than peak hour or sometimes a bit high. This condition is true for all types of cars, buses, taxicabs, etc. For the entire Dhaka city, the gross working hours saving are described in the following Table 4.11. So, around .2 million working hours per day using bicycle in the motijheel area.

**Time Saving Explanation:**

Total Bus ride reduce: $0.1 \times 127398 \times 20$ km $= 254796$ km

Time saving for Bus: $254796 \times 40 \times (8.57 - 1.5)$ mins $= 7.2 \times 10^7$ mins

$\qquad\qquad = 1.2 \times 10^6$ hours $= 50000$ day

Total Car ride reduce: $0.05 \times 293268 \times 10$ km $= 146634$ km

Time saving for Car: $146634 \times (8.57 - 1.5)$ mins $= 1.04 \times 10^6$ mins

$\qquad\qquad = 1.73 \times 10^4$ hours $= 720$ days

Total Taxicab ride reduce: $0.2 \times 36600 \times 100$ km $= 732000$ km

Time saving for Taxicab: $732000 \times (8.57 - 1.5)$ mins $= 5.2 \times 10^6$ mins

$\qquad\qquad = 8.66 \times 10^4$ hours $= 3611$ days

Total Motorcycle ride reduce: $0.5 \times 724800 \times 15$ km $= 5.4 \times 10^6$ km

Time saving for Motorcycle: $5.4 \times 10^6 \times (8.57 - 1.5)$ mins $= 3.8 \times 10^7$ mins

$\qquad\qquad = 6.3 \times 10^5$ hours $= 26289$ days

Total working time saving: $50000 + 720 + 3611 + 26289$ days $= 80620$ days

### 4.3.2.2 Fuel and Cost Saving

The cost of installation or repair is not listed here rather we considering only running cost. Because whenever users switch from car to bicycle there would be a significant reduction in costs.

Here it is assumed that the mileage is 5km per liter diesel for bus and 20km per liter diesel for both car and taxicab of 65tk per liter and the mileage is estimated to be 50km per liter for motorcycle of 89tk per liter. On the other hand, the electric cycle charges of 10tk with 50km of travel per charge charges. The fuel consumption is calculated per day basis and user saving is grand saving of considering all users of the Dhaka city.

In the following Table 4.12, the needed fuel for cars, taxicabs, motorcycles, and buses is calculated and there is no running cost & fuel cost for the bicycle. In case of bus, the per km fare is fixed by BRTA of 1.7tk and we assumed that taxicab fare is 50tk per km. Here the distance of all the node point is measured from center node point 7.

Table 4.12: Cost comparison in car, bus and bicycle considering from node point 7.

| Node Point | Distance (km) | Car Fuel (litre) | Car User Cost (tk) | Taxicabs Fuel (litre) | Taxicabs User Cost (tk) | Motor Cycle Fuel (litre) | Motor Cycle User Cost (tk) | Bus Fuel (litre) | Bus User Cost (tk) | Electric Bicycle Distance (km) | Electric Bicycle User Cost (tk) |
|---|---|---|---|---|---|---|---|---|---|---|---|
| 1 | 14 | 0.7 | 45.5 | 0.56 | 700 | 0.28 | 15.49 | | 23.8 | 15.65 | 3.13 |
| 2 | 7.5 | 0.38 | 24.38 | 0.3 | 375 | 0.15 | 10.68 | | 12.75 | 9.89 | 1.98 |
| 3 | 18.2 | 0.91 | 59.15 | 0.73 | 910 | 0.36 | 8.37 | | 30.94 | 19.51 | 3.9 |
| 4 | 10.2 | 0.51 | 33.15 | 0.41 | 510 | 0.2 | 6.23 | | 17.34 | 9.02 | 1.8 |
| 5 | 8.7 | 0.44 | 28.28 | 0.35 | 435 | 0.17 | 9.79 | | 14.79 | 11.9 | 2.38 |
| 6 | 6 | 0.3 | 19.5 | 0.24 | 300 | 0.12 | 16.2 | As we assume bicycles reduces 10% of the total bus in Dhaka city. The fuel consumption is reduced = 12740×20 ×1/5 litres = 50959 litres | 10.2 | 3.95 | 0.79 |
| 8 | 4.7 | 0.24 | 15.28 | 0.19 | 235 | 0.09 | 21.36 | | 7.99 | 3.17 | 0.63 |
| 9 | 3.5 | 0.18 | 11.38 | 0.14 | 175 | 0.07 | 17.27 | | 5.95 | 2.17 | 0.43 |
| 10 | 5.5 | 0.28 | 17.88 | 0.22 | 275 | 0.11 | 14.77 | | 9.35 | 5.98 | 1.2 |
| 11 | 9.1 | 0.46 | 29.58 | 0.36 | 455 | 0.18 | 15.49 | | 15.47 | 10.61 | 2.12 |
| 12 | 12 | 0.6 | 39 | 0.48 | 600 | 0.24 | 14.24 | | 20.4 | 9.04 | 1.81 |
| 13 | 9.7 | 0.49 | 31.53 | 0.39 | 485 | 0.19 | 29.55 | | 16.49 | 9.44 | 1.89 |
| 14 | 8.3 | 0.42 | 26.98 | 0.33 | 415 | 0.17 | 20.29 | | 14.11 | 7.31 | 1.46 |
| 15 | 8.7 | 0.44 | 28.28 | 0.35 | 435 | 0.17 | 15.49 | | 14.79 | 11.55 | 2.31 |
| 16 | 8 | 0.4 | 26 | 0.32 | 400 | 0.16 | 3.92 | | 13.6 | 8.7 | 1.74 |
| 17 | 16.6 | 0.83 | 53.95 | 0.66 | 830 | 0.33 | 4.09 | | 28.22 | 15.61 | 3.12 |
| 18 | 11.4 | 0.57 | 37.05 | 0.46 | 570 | 0.23 | 12.1 | | 19.38 | 14.78 | 2.96 |
| 19 | 8.7 | 0.44 | 28.28 | 0.35 | 435 | 0.17 | 11.75 | | 14.79 | 9.87 | 1.97 |
| 20 | 2.2 | 0.11 | 7.15 | 0.09 | 110 | 0.04 | 15.49 | | 3.74 | 2.12 | 0.42 |
| 21 | 2.3 | 0.12 | 7.48 | 0.09 | 115 | 0.05 | 28.3 | | 3.91 | 1.76 | 0.35 |
| 22 | 6.8 | 0.34 | 22.1 | 0.27 | 340 | 0.14 | 26.17 | | 11.56 | 7.34 | 1.47 |
| 23 | 6.6 | 0.33 | 21.45 | 0.26 | 330 | 0.13 | 15.13 | | 11.22 | 4.07 | 0.81 |
| 24 | 8.7 | 0.44 | 28.28 | 0.35 | 435 | 0.17 | 18.16 | | 14.79 | 8.55 | 1.71 |
| 25 | 15.9 | 0.8 | 51.68 | 0.64 | 795 | 0.32 | 34 | | 27.03 | 18.72 | 3.74 |
| 26 | 14.7 | 0.74 | 47.78 | 0.59 | 735 | 0.29 | 15.49 | | 24.99 | 15.5 | 3.1 |
| 27 | 8.5 | 0.43 | 27.63 | 0.34 | 425 | 0.17 | 10.68 | | 14.45 | 11.2 | 2.24 |
| 28 | 10.2 | 0.51 | 33.15 | 0.41 | 510 | 0.2 | 8.37 | | 17.34 | 9.41 | 1.88 |
| 29 | 19.1 | 0.96 | 62.08 | 0.76 | 955 | 0.38 | 6.23 | | 32.47 | 22.02 | 4.4 |

Table 4.13: Cost saving per day.

| Transit | % of Transit Reduction | Transit Reduces | Riding Distance (km) | Fuel (litres) | User Cost (tk) |
|---|---|---|---|---|---|
| Bus | 10% | 12740 | 254796 | 50959 | 1.27 million |
| Car | 5% | 14663 | 146630 | 7331 | 4.5 lakh |
| Taxicab | 20% | 7320 | 732000 | 29280 | 36.5 million |
| Motor cycle | 50% | 362400 | 5436000 | $1.08 \times 10^5$ | 8.5 million |
| | | | Total cost saving | 195570 | 46.72 million |

The gross fuel saving and user cost saving of the entire Dhaka city is described in the following Table 4.13. So, it is saved around 195570 liters fuel and 46.72 million user costs per day for using bicycle in the entire Dhaka area.

**Fuel and Cost Saving Explanation:**

Fuel saving for Bus: 254796 km × 1/5 litres = 50959 litres

User money saving: $254796 \text{ km} \times (\frac{65}{5} - 40 \times \frac{10}{50})$ tk = 1.27 million tk

Fuel saving for Car: $146634 \text{ km} \times \frac{1}{20}$ litres = 7331 litres

User money saving: $146634 \text{ km} \times (\frac{65}{20} - \frac{10}{50})$ tk = $4.5 \times 10^5$ tk = 4.5 lakh tk

Fuel saving for Taxicab: $732000 \text{ km} \times \frac{1}{25}$ litres = 29280 litres

User money saving: $732000 \text{ km} \times (50 - \frac{10}{50})$ tk = $3.65 \times 10^7$ tk = 36.5 million tk

Fuel saving for Motor cycle: $5.4 \times 10^6 \text{ km} \times \frac{1}{50}$ litres = $1.1 \times 10^5$ litres

User money saving $5.4 \times 106 \times (\frac{89}{50} - \frac{10}{50})$ tk = $8.5 \times 10^6$ tk = 8.5 million tk

Total fuel saving: $50959 + 7331 + 29280 + 1.1 \times 10^5$ litres = 197570 litres

Total User money saving: $1.27 + .45 + 36.5 + 8.5$ million tk = 46.72 million tk

### 4.3.2.3 $CO_2$ Emission Reduction

In the following calculation, 887 g/km, 258 g/km, 237 g/km and 40 g/km are the considered amount of $CO_2$ emission in 1km ride of bus, car, taxicab, and motorcycle respectively. Then $CO_2$ emission is reduced in a significant amount. For the entire Dhaka city, the gross $CO_2$ emission reduction is described in the Table 4.14. In total, around $6.58 \times 10^5$ kg $CO_2$ emission is reduced in the entire Dhaka city per day.

Table 4.14: $CO_2$ emission reduction per day.

| Transit | % of Transit Reduction | Transit Reduces | Riding Distance (km) | $CO_2$ Emission (g) |
|---|---|---|---|---|
| Bus | 10% | 12740 | 254796 | $2.3 \times 10^8$ |
| Car | 5% | 14663 | 146630 | $3.8 \times 10^7$ |
| Taxicab | 20% | 7320 | 732000 | $1.7 \times 10^8$ |
| Motor cycle | 50% | 362400 | 5436000 | $2.2 \times 10^8$ |
| | | | Total $CO_2$ emission reduction | $6.58 \times 10^8$ |

**$CO_2$ Emission Reduction Explanation:**

$CO_2$ emission reduction for Bus: $0.1 \times 127398 \times 20 \text{ km} \times 887 \text{ gm} = 2.3 \times 10^5$ kg

$CO_2$ emission reduction for Car: $0.05 \times 293268 \times 10 \text{ km} \times 258 \text{ gm} = 3.8 \times 10^4$ kg

$CO_2$ emission reduction for Taxicab: $0.2 \times 36600 \times 100 \text{ km} \times 237 \text{ gm} = 1.7 \times 10^5$ kg

$CO_2$ emission reduction for Motorcycle: $0.5 \times 724800 \times 15 \text{ km} \times 40 \text{ gm} = 2.2 \times 10^5$ kg

Total $CO_2$ emission reduction: $6.58 \times 10^5$ kg

When CNG is being used as fuel in buses, cars, and taxicabs the $CO_2$ emission is more significant. Buying a car and motorcycle also increases traffic jams, air carbon dioxide, pollution of the environment and costly in the current situation in Dhaka. Alternatively, electric bicycle does not make an enormous traffic jam, $CO_2$ in air, and less expensive. In order to do this, governments should build parking lots in various places where necessary.

# CHAPTER 5

# Conclusions

A modified Physarum-inspired model is presented in this paper to address the design of the bicycle lane network. Different approaches, like exact approaches and heuristic approaches, have been presented over the past decades to design transportation networks. Recently bio-inspired method had drawn great attraction to network design. In real two-way traffic networks, the modified technique is more effective and efficient. This chapter will now give a short summary of the main points described in this thesis. Also, it discusses possible future works based on the outcome of the present work

## 5.1 Achievements

The network design technology inspired by Physarum is believed to have balanced costs, effectiveness, and resilience. Inside Dhaka city, an unorganized and unplanned city, we have developed an electric bicycle network system where there's little footway. To meet this challenge, we primarily use local roads and try to avoid major roads towards the construction of the electric bicycle network. Since bicycles are non-motorized vehicles do not produce greenhouse gases so they do not cause air pollution. They also don't contribute to noise pollution. If a large number of people use bicycle in the city, traffic jams will be eliminated. The costs will be reduced and people can have some physical activity also, which is beneficial to health. Since bicycles do not need to use gasoline, the importation of gasoline will be reduced. That also enriches the economy and the environment.

## 5.2 Future Study

In the future, parallel computing and the optimal model for the design of the transport network are part of our work. Furthermore, our research includes the implementation of the Physarum polycephalum inspired model for the dynamic traffic network and the elastic demand traffic network.